\documentstyle[12pt,epsfig]{article}
 1
\def\as{\alpha_{\rm S}}

\def\citenum#1{{\def\@cite##1##2{##1}\cite{#1}}}
\def\citea#1{\@cite{#1}{}}

\def\as{\alpha_{\rm S}}

\def\b{\beta}
\def\a{\alpha}

\def\D{\Delta}

\def\g{\gamma}
\def\G{\Gamma}

\def\l{\lambda}

\def\o{\omega}

\def\pa{\partial}

\def\ra{\rightarrow}

\def\S{\Sigma}
\def\s{\sigma}

\def\ti{\tilde}

\def\({\left(}
\def\){\right)}

\def\citenum#1{{\def\@cite##1##2{##1}\cite{#1}}}
\def\citea#1{\@cite{#1}{}}

\def\l1vt{\vec{l_{1\perp}}}

\def\bt{b_{\perp}}

\def\bt2{$b^2_t$}
\def\aa{$z(1 - z)\, Q^2\,\,+\,\,m^2_Q$}

\def\jol1{$J_0(\,l_{1\perp}\,r_{\perp}\,)$}
\def\ko{K_0(\,a\,r_{\perp}\,)}

\def\citea#1{\@cite{#1}{}}

\def\beq{\begin{equation}}
\def\eeq{\end{equation}}
\def\bea{\begin{eqnarray}}
\def\eea{\end{eqnarray}}

\def\eq#1{{eq.~(\ref{#1})}}

\def\bbbz{{\mathchoice {\hbox{$\sf\textstyle Z\kern-0.4em Z$}}
{\hbox{$\sf\textstyle Z\kern-0.4em Z$}}
{\hbox{$\sf\scriptstyle Z\kern-0.3em Z$}}
{\hbox{$\sf\scriptscriptstyle Z\kern-0.2em Z$}}}}
\def\npb#1#2#3{    {\it Nucl. Phys. }{\bf B#1} (19#2) #3}
\def\plb#1#2#3{    {\it Phys. Lett. }{\bf B#1} (19#2) #3}
\def\prd#1#2#3{    {\it Phys. Rev. }{\bf D#1} (19#2) #3}

\def\zpc#1#2#3{    {\it Z. Phys. }{\bf C#1} (19#2) #3}

\def\sjnp#1#2#3{   {\it Sov. J. Nucl. Phys. }{\bf #1} (19#2) #3}

\relax
\begin{document}
\begin{titlepage}
\noindent
\begin{flushright}
 May  1996 \\ CBPF-NF-021/96 \\
TAUP 233896/96
\end{flushright}
\begin{center}
{\Large\bf DIFFRACTIVE LEPTOPRODUCTION     }   \\[2.4ex]
{\Large \bf OF SMALL MASSES IN QCD  }\\[11ex]
{\large E R R O L\,\, G O T S M A N}  $^{1)}$
\footnotetext{$^{1)}$ Email: gotsman@post.tau.ac.il .}\\[1.5ex]
{\it School of Physics and Astronomy}\\
{\it Raymond and Beverly Sackler Faculty of Exact Science}\\
{\it Tel Aviv University, Tel Aviv, 69978, ISRAEL}\\[3.5ex]
{\large E U G E N E \,\, L E V I N}  $^{2)} $
\footnotetext{$^{2)}$ Email: levin@fnalv.fnal.gov .}\\[1.5ex]
{\it  LAFEX, Centro Brasileiro de Pesquisas F\'\i sicas  (CNPq)}\\
{\it Rua Dr. Xavier Sigaud 150, 22290 - 180 Rio de Janeiro, RJ, BRASIL}
\\{\it and}\\
{\it Theory Department, Petersburg Nuclear Physics Institute}\\
{\it 188350, Gatchina, St. Petersburg, RUSSIA}\\[3.5ex]
{\large U R I \,\,M A O R}  $^{3)}$
\footnotetext{$^{3)}$ Email: maor@ccsg.tau.ac.il .}\\[1.5ex]
{\it School of Physics and Astronomy}\\
{\it Raymond and Beverly Sackler Faculty of Exact Science}\\
{\it Tel Aviv University, Tel Aviv, 69978, ISRAEL}\\[7ex]
\end{center}
\newpage
~\,\,\,
\vspace{4cm}

{\large \bf Abstract:}
In this paper we consider the process of diffraction dissociation in
deep inelastic scattering producing  a small  mass.
This process is analyzed by the calculation of
$\bar q\, q$ and $\bar q\, q\, G$ production.
We show that the small distance contributions ($r_{\perp}\,\propto\,1/Q$)
to  the longitudinal polarised virtual
photon dominate the diffractive channels. Formulae for the cross section
using the gluon density are written within the framework of
perturbative QCD. We  show that the production of small masses by
a transverse polarised photon is concentrated at moderate values of
$r_{\perp} \,\sim\,1 GeV^{-1}$, where the pQCD approach is valid.
It is shown that only $\bar q\, q$ pair production contributes to
diffraction dissociation with $\b\,>\,$ 0.4 and the possibility
to extract the values of the
gluon density from the measurements in this kinematic
region is discussed.
Shadowing corrections are assessed for both longitudinal and
transverse polarised photons
and estimates of the different damping factors are given. The relation
between diffractive production and the corrections to $F_2$ is alluded
to. The evolution of the diffractive structure function is studied and
a solution to the diffractive evolution equations is proposed.

\end{titlepage}

\section{Introduction}
\par
Over the last decade, starting with the paper of Bartels and Loewe\cite
{BAL},
diffractive processes in deep inelastic scattering (DIS)
have attracted a great deal of attention.
The reason is that these processes can be
calculated within the framework of
perturbative QCD (pQCD)\cite{DD} and their calculated
cross sections turn out to be proportional
to the square of the gluon density
$x_P G(Q^2,x_P)$.
$Q^2$ is
the virtuality of the photon and $x_P = \frac{Q^2 + M^2}{s}$, where
$s$ is the c.m. energy squared of the reaction. $M$
is the mass of the diffractively produced hadronic
system. The recent renewed interest in these channels was triggered
by Ryskin\cite{RY} and Brodsky et.al.\cite{FIVE}, who
proved that vector meson electro production
can be  calculated in the leading
log approximation of pQCD, while the non perturbative effects,
originating from large distances, can be factored out in terms of
the light-cone $\bar q q$  wave function of the produced
vector meson.
\par
The goal of this paper is to study single diffraction dissociation
(DD) production of a small mass hadronic system.
We consider a small mass as  the hadronized product
 of either a $\bar q q $ pair or a $\bar q q G $ system in the
final state
of the DD process. We will show that:
\newline
1. This process
describes a significant part of the experimental DD cross section.
\newline
2. The small distance contribution ($ r_{\perp} \,\propto\,1/Q $)
dominates the longitudinal polarised virtual photon
which  induces DD. This justifies the use of a
pQCD approach
to calculate the relevant cross section.
\newline
3. The production of small masses by a transverse polarised photon
are concentrated at moderate values of
$r^2_{\perp}\,\,\approx\,\, 1 \, GeV^{-2}$. This encourages us
to believe, that our estimates which are obtained
in pQCD, could be responsible for a
considerable part of the DD cross section.
\newline
4. Shadowing (or screening) corrections (SC) are not negligble in the
region of low $x_P$ and
should be taken into account.
\par
The plan of our paper is as follows:
In the next section we calculate the process of DD
by longitudinal and transverse
polarized photons into a $\bar q q $ pair
as well as through $\bar q q G$.
This is done in the $r_{\perp}$-representation, which provides a
natural frame
to assess at what distances these processes are viable. It is also
a convenient representation to calculate SC. Most of the formulae
which we obtain for transverse polarised photons
have been  previously derived in the momentum representation, see
Refs.\cite{NZ1}\cite{RYS}\cite{LW}\footnote
{Mark W\"{u}sthoff has done the calculations
for a longidudinal polarized photon in his PhD thesis and presented the
results in the 1995 Durham workshop.}.
In section 3 we extend the formalism suggested
by Levin and Ryskin\cite{LR} and Mueller\cite{M90}, to the case of a
penetration of a $\bar q q $ pair with a definite value of $r_{\perp}$
through the proton,
as well as through the nucleus. We will show that for the case of a
diffractive production of small masses
by longitudinal polarised photons,
we obtain an elegant closed expression for the
damping factor\cite{GLMD} defined by
\beq \label{1}
D^2\,\,=\,\,\frac{\frac{d \s(\g^* + p \ra X(M) + p )}{d t} [with \,
\,SC]}
{\frac{d \s( \g^* + p \ra X(M) + p)}{d t}[without\,\,SC]}\,\mid_
{t = 0}
\eeq
where $ t = - q^2_{\perp}$.
Processes at large transverse distances do not enter the above formula,
allowing
us to calculate the SC within the framework of pQCD.
We extend our formalism also to DD
by a transverse
polarized photon.
We  discuss the large distance contribution in this case and show that
the SC lead to a decrease of the typical distances that are responsible
for DD.
In section 4 we compare the pQCD predictions with
the experimental data available from
HERA\cite{DDH1}\cite{DDZEUS}
(see also Refs.\cite{ZEUSNEW}\cite{H1DDNEW}).
We show that pQCD is able to account for a considerable part of the
observed
DD cross section, and discuss which distances contribute to the different
modes present in the final state of the produced system.
In all our calculations we use the GRV parametrization\cite{GRV}
for the gluon density, which is
the main input ingredient in our work.
The GRV parametrization which starts from a low value
of the photon virtuality, allows us to study the value of typical
distances in the DD process.
In the fifth section we discuss
the evolution equations for the DD processes.
A summary of our results and a discussion are presented in
the conclusions.

\section{The diffractive production of {\bf$\bar q\, q$} and {\bf
$ \bar q\, q\, G$} systems without shadowing corrections}
\subsection{Notation}
\par
We list below the notation used in this paper
(see Fig.1).
\newline
1.
$Q^2$ is the virtuality of the photon in DIS.
\newline
2.
$x_{P} \,\,=\,\,\frac{Q^2 + M^2}{s}$, where $s$ denotes the square of the
c.m. energy
and $M$ is the produced diffractive mass.
Bjorken's scaling variable is given by $ x_B\,\,=\,\,\frac{Q^2}{s}$.
\newline
3.
$\b\,\,=\,\,\frac{Q^2}{Q^2\,\,+\,\,M^2}\,\,=\,\,\frac{x_B}{x_{P}}$.
\newline
4.
$ a^2 $= \aa, where $z$  is the fraction of the photon energy that is
carried away by a quark with a mass $m_Q$.
\newline
5.
$\vec{k}_{\perp}$ denotes  the transverse momentum of the quark, and
$\vec{r}_{\perp}$  the transverse
distance between the quark and the antiquark.
\newline
6.
$\vec{b}_{\perp}$ is the impact parameter
 of the reaction and is the conjugate variable to
$\vec{q}_{\perp}$, the momentum transfer from the
incoming  proton to the recoiled proton. Note that
$t = - q^2_{\perp}$.
\newline
7.
$\vec{l}_{i\perp}$ denotes the transverse momentum of gluons,
 which are  attached
to the quark-antiquark
pair (see Fig.1).
\newline
8.
We use the evolution equations for the parton densities
in moment space. For any function $f(x)$, we  define the moment $f(\o)$
as
\beq \label{2}
f (\o)\,\,=\,\,\int^1_0\,\,dx \,\,x^{\o}\,\,f(x)\,\,\,\,
\eeq
Note that the moment variable $\o$ is chosen so that the $\o$ = 0 moment
measures the number of partons and the $\o$ = 1 moment measures their
momentum. An alternative moment variable N, defined such that N = $\o$ +
1,
is often found in the literature.
\par
The $x$ - distribution can be reconstructed by considering the inverse
Mellin transform. For example, for the gluon density it reads
\beq \label{3}
x G(Q^2,x)\,\,=\,\,\frac{1}{2 \pi i}\,\,\int_C \,\, d \,\o \,\,x^{- \o}
\,\,g (Q^2,\o)
\eeq
The contour of integration C is taken to the right of all singularities.
\par
For the solution of the DGLAP equations\cite{GLAP},
as well as the BFKL equation\cite{BFKL}, one
has the following form
\beq \label{4}
g(Q^2,\o)\,\,=\,\,g(\o)\,e^{\g(\o) ln Q^2}
\eeq
where $\g(\o)$ denotes the anomalous dimension which, in the leading
log $\frac{1}{x}$ approximation of pQCD, is
a function of $\frac{\as}{\o}$ and can be
presented as\cite{LIANDI}
\beq \label{5}
\g ( \omega ) \,\,=\,\,\frac{\as N_C }{\pi} \frac{1}{\omega} \,\,+\,\,
\frac{2  \as^4 N^4_C \zeta ( 3 )}{\pi^4} \frac{1}{\omega^4}\,\,+\,\,O
( \frac{\as^5}{ \omega^5} )
\eeq
9.
Our amplitude is normalized so that
$$
\frac{d \s}{d t}\,\,=\,\,\pi \,|f(s,t)|^2
$$
and the optical theorem can be written
$$
\s_{tot}\,\,=\,\,4\pi\,\,Im\,f (s,0)
$$
The scattering amplitude in $b_{\perp}$-space is defined by
$$
a(s,b_{\perp})\,\,=\,\,\frac{1}{2 \pi}\,\,\int\,\,d^2 q_{\perp} \,\,
e^{- i {\vec{q}}_{\perp}\,\cdot\,{\vec{b}}_{\perp}}\,\,f(s,t = - q^2_
{\perp})$$
In this representation
$$\s_{tot}\,\,=\,\,2\,\,\int \,d^2 b_{\perp} \,\,Im\,\,a (s, b_{\perp} )
$$
and
$$
\s_{el}\,\,=\,\,\int \,d^2 b_{\perp} \,\,|\,a (s, b_{\perp} )|^2
$$
\newline
In  what follows we use the notation and normalization
of Brodsky et.al.\cite{FIVE}.

\subsection{The cross section for diffractive dissociation}
\par
The cross section for the diffractive process
(see Fig.1)
\beq \label{6}
\g^* (Q^2,x_B) \,\,\,+\,\,\,p\,\,\,\ra\,\,\,X\,( M^2 )\,\,\,+\,\,p
\eeq
can be written in the form
\beq \label{7}
x_{P}\,\frac{d \s}{ d x_P d t}\,\,=\,\,\frac{3 \alpha_{em}}
{ 8 (2 \pi)^2}\,\sum^{N_F}_{1}\,Z^2_F\,\, \sum_{\lambda_1 \lambda_2}
\int^1_0 \,\,d\,z\,\,\int\,\,
\frac{ d\,k^2_{\perp}}{ (2 \pi)^2}
\eeq
$$
\,\,\, | f (\, \g^*\,+\,p\,\ra \,
q (z, k_{\perp}) \,+\,\bar q (1 - z, - k_{\perp} \,+\, q_{\perp})\,\,
+\,\,p\,)|^2\, \,(\,M^2\,+\,Q^2\,)\,\,\delta \(\, M^2 \,-\,
\frac{k^2_{\perp}}{z ( 1 - z )}\,\)
$$
We have assumed in eq.(7) that the small masses are produced by
the dissociation
of a $\bar q q $ system. We will consider later,
 the corrections resulting from  the emission of
gluons. In eq.(7), $\lambda_i$ denotes the polarisation of the quark,
$N_F$  the number of quark flavours and $Z_F$ the fraction of the
electron charge carried by the quark with a flavour $F$.
\par
Integrating over $z$  and summing over polarizations,
we reduce the above equation to the form
\beq \label{8}
x_{P}\,\frac{d \s}{ d x_P d t}\,\,=\,\,\,\frac{3 \alpha_{em}}
{(2 \pi)^3}\,\sum^{N_F}_{1}\,Z^2_F\,\,\,N_{\lambda}
\,\int\,\, | f|^2\,\,\frac{d^2 \,k_{\perp}}{1\,-\,\b}\,
 \,\frac{k^2_{\perp}}{M^2}\,\frac{2}{\sqrt{1\,-\,\frac{4 k^2_{\perp}}
{M^2}}}
\eeq
where $z ( 1 - z) \,=\,\frac{k^2_{\perp}}{M^2}$ and $a$\, =\,$\sqrt{z(1
- z) Q^2 + m^2_Q}$\, =\,
$Q\,\frac{k_{\perp}}{M}$ for massless quarks. The region of integration
with respect to $k_{\perp}$ is defined $ k^2_{\perp}\,\,<\,\,\frac{M^2}
{4}$.
The factor $N_{\lambda}$ depends on the initial polarization of the
photon.
In the case of longitudital polarization $N_{\lambda} $= 4. For
transverse polarization $N_{\lambda}\, = \,z^2\, + \,(1 - z )^2$.
We follow Refs.\cite{NZ1}\cite{M90} regarding normalization, and the
definition of the photon wave function.
The main task before us is the calculation of the
amplitude $f$.

\subsection{The amplitude in the $r_{\perp}$ representation}
\par
This approach was first formulated in Ref.\cite{LR} and
has been carefully developed in Ref.\cite{M90}.
During it's time of passage
through the target, the distance $r_{\perp}$ between a quark and
an antiquark
can vary by an amount $\D r_{\perp}\,\,\propto R \,\,\frac{k_{\perp}}{E}
$.
$E$ denotes the energy of the pair and $R$ the size of the target
(see Fig.1).
The quark's transverse
momentum is $k_{\perp} \,\propto \,\frac{1}{r_{\perp}}$. Therefore
\beq \label{9}
\D r_{\perp}\,\,\propto\,\,R \,\frac{k_{\perp}}{E}\,\,\ll\,\,r_{\perp}
\eeq
holds if
\beq \label{10}
r^2_{\perp} s \,\gg 2 m \,R
\eeq
where $s = 2m E $.
\par
The above condition can be rewritten in terms of $x_P$
\beq \label{11}
x_P\,\,\ll\,\,\frac{2}{ ( 1 - \b)\, m R }
\eeq
This means that at small values of $x_P$, the transverse distance
between the quark and the
antiquark is a good degree of freedom\cite{LR}\cite{M90}\cite{MU}, and
the interaction of a virtual photon with the target can be written in
the form
\beq \label{12}
f\,\,= \,\,\int\,\,\frac{d^2 r_{\perp}}{2 \pi}\,\,\,
{\Psi}^* (r_{\perp},k_{\perp})
\,\,\s(r_{\perp},q^2_{\perp})\,\, \Psi^{\g^*}(Q^2,r_{\perp},k_{\perp})
\eeq
$\Psi$ denotes the wave function of the produced $\bar q q$ pair which
is equal to
$e^{i \vec{r}_{\perp}\,\cdot\,\vec{k}_{\perp}}$. After integrating
over the azimuthal angle we have
$J_0( \,k_{\perp}\,r_{\perp}\,)$  and $J_1( \,k_{\perp}\,r_{\perp}\,)$
for the longitudinal and transverse polarised photon induced reactions
respectively.
$\Psi^{\g^*}$ denotes the wave function of the longitudinally ( L )
or transverse ( T ) polarized photon.
These have been given in Refs.\cite{NZ1}\cite{M90} and are equal to
\beq  \label{13}
\Psi^{\gamma^*}_L(Q^2,r_{\perp},z)\,\,=\,\,
  Q z ( 1 - z) \ko\,\,= \,\,
\,\,Q \,\frac{k^2_{\perp}}
{M^2}\,\,\ko
\eeq
and
\beq \label{57}
\Psi^{\gamma^*}_{T}(Q^2,r_{\perp},k_{\perp})\,\,=\,\,i\,a\,
K_1(a \,r_{\perp})\,\,\frac{\vec{r_{\perp}}}{r_{\perp}}\,\,
\eeq
$\s(r_{\perp},q^2_{\perp})$ denotes the cross section of the $\bar q q $
pair
with a transverse separation $r_{\perp}$, which scatters with transverse
momentum $q_{\perp}$. We will discuss it in detail below,
considering initially
the case where  $q_{\perp}$ = 0.

\subsection{$\s(r_{\perp},q^2_{\perp})$ at $q_{\perp}$ = 0}
\par
The form for
$\s(r_{\perp},q^2_{\perp})$ at $q_{\perp}$ has been given in Ref.\cite
{LR}
(see \eq{8} of this paper).  $\s$ can be expressed in terms of the
unintegrated parton density $\phi$, first introduced in the BFKL
papers\cite
{BFKL} and widely used in Ref.\cite{GLR}. This function is clearly
related to the gluon density and can
be calculated using the equation
\beq \label{14}
\as (Q^2) x G(Q^2,x)\,\,=\,\,\int^{Q^2} \,\,d l^2_{\perp}\,\,
\as(l^2_{\perp})\,\,
\phi(l^2_{\perp}, x )
\eeq
Using the above equation we obtain the result of Ref.\cite{LR},
which reads
\beq \label{15}
\s(r_{\perp},q^2_{\perp})\,\,=\,\,\frac{16 C_F  }{N^2_C - 1}\,\,\pi^2
\int \phi(l^2_{\perp},x)\,\,(\,\,1\,\,-\,\,e^{i\,\vec{l}_{\perp}\cdot
\vec{r}_{\perp}}\,\,)\,\,\,\frac{\as(l^2_{\perp})}{2 \pi}\,\,
\frac{d^2 l_{\perp}}{l^2_{\perp}}
\eeq
where $\phi \,\,=\,\,\frac{\pa x G (Q^2,x)}{\pa  Q^2}$.
\par
We calculate this integral using eqs.(3) and (4), and integrate over the
azimuthal angle. Introducing a new variable $\xi = r_{\perp} l_{\perp}$
we can reduce the integral to the form
\beq \label{16}
\s(r_{\perp},q^2_{\perp})\,\,=\,\,\frac{16 C_F \as }{N^2_C - 1}\,\,\pi^2
\int_C \frac{d \o}{ 2 \pi i} \,\,g(\o)\,\,\g(\o)\,\,(r^2_{\perp})^{1 -
\g(\o)}
\int^{\infty}_{0} d \,\,\xi\,\, \frac{1 - J_0(\xi)}{ (\xi)^{3 - 2
\g(\o)}}
\eeq
preforming the integration over d$\xi$
(see Ref.\cite{AB} {\bf 11.4.18}) we have
\beq \label{17}
\s(r_{\perp},q^2_{\perp})\,\,=\,\,\frac{8 C_F \as }{N^2_C - 1}\,\,\pi^2
\int_C \frac{d \o}{ 2 \pi i} \,\,g(\o)\,\,\g(\o) \,\,
(\frac{r^2_{\perp}}{4})^{1 - \g(\o)}\,\,\,\,\frac{\G(\g(\o)) \,\,
\G(1 - \g(\o))}{(\,\G(2 - \g(\o)\,)\,)^2}
\eeq
In the double leading log approximation of pQCD, $\g(\o)\,\,\ll\,\,1$,
and we obtain
the following cross section for $N_C$ = 3
\beq \label{18}
\s(r_{\perp},q^2_{\perp})\,\,=\,\,\frac{\as(\frac{4}{r^2_{\perp}})}{3}
\,\,\pi^2\,\,r^2_{\perp}\,\,
\(\,\, x G^{DGLAP}\(\frac{4}{r^2_{\perp}},x\)\,\,\)^2
\eeq
This result agrees with the expression for the cross section value given
in Refs.\cite{KNNZ} and
\cite{AFS} provided we neglect the factor 4 in the argument of the gluon
density. We have checked that eq. (2.16) of Ref.\cite
{FIVE} also leads to the
same result, unlike the value for $\s$ quoted in the paper (see
eq. (2.20) in Ref.\cite{FIVE}).

\subsection{The amplitude for small mass production at $q_{\perp}$ = 0
induced by a longitudinal photon}
In this section we calculate the amplitude for
$\bar q q $ production with a mass $M$. Substituting
\eq{13}, in our expression (12) for
$f$, yields
\beq \label{19}
f\,\,=\,\,\int\,\, r_{\perp} d r_{\perp}\,\,\,
\frac{Q\,k^2_{\perp}}{M^2}\,K_0(\,\frac{Q}{M} \,k_{\perp} \,r_{\perp}\,)
\,\,\frac{8 C_F \as }{N^2_C - 1}\,\,\pi^2
\int_C \frac{d \o}{ 2 \pi i} \,
\eeq
$$
\,g(\o)\,\,\g(\o) \,\,
(\frac{r^2_{\perp}}{4})^{1 - \g(\o)}\,\,\,\,\frac{\G(\g(\o)) \,\,
\G(1 - \g(\o))}{(\,\G(2 - \g(\o)\,)\,)^2}
\,\, J_0 (k_{\perp}\,\,r_{\perp})
$$
Making use of the well known properties of the modified Bessel functions
we can perform the integration.
The final general result is
\beq \label{20}
f\,\,=\,\,\,
 \frac{Q k^2_{\perp}}{M^2}\frac{8C_F \as}{N^2_C - 1}\,\,\pi^2
\int_C \frac{d \o}{ 2 \pi i} \,\,g(\o)\,
\,\, \frac{1}{a^4}\,\,
(\frac{a^2}{\b})^{\g(\o)}\,\,
\eeq
$$
\,\,\G\(1 + \g(\o)\) \,\,
\G\(1 - \g(\o)\) \,\G\(2 - \g(\o)\)\,\G\(2 - \g(\o)\)
\,\b^3 \,\,{}_1F_1 \(\, 2 \,-\,\g(\o), -1\,+\,\g(\o),
 1, \,1\,- \,\b\,\)
$$
where $a \,=\,Q\,\frac{k_{\perp}}{M}$.
\par
In the case of the DGLAP approach we can simplify \eq{20} as
$\g(\o) \,\,\ll\,\,1$. Neglecting $\g(\o)$ with respect to 1,
and using \eq{3} we get
\beq \label{21}
f\,\,=\,\,-\,\pi^2\,\,\frac{8\,C_F\, \as}{N^2_C - 1}\,\,
\frac{Q \,\frac{k^2_{\perp}}{M^2}}
{(\,Q^2 \,\frac{k^2_{\perp}}{M^2}\,)^2}\,\b^2\,x_P\,
G\(\,\frac{k^2_{\perp}}{\,( 1 - \b )},x_P\,\)\,\,(\, 1 \,-\,2 \,\b\,)
\eeq
It is interesting to note that the argument of $x_P G $ is
$k^2_{\perp}/ ( 1 - \b)$, which means that small distances of the order
of $r^2_{\perp} \,\,\propto \,\,( 1 - \b )/ k^2_{\perp}$, contribute to
small mass diffractive production, especially at $\b \,\rightarrow \,1 $.

\subsection{DD cross section induced by a longitudinal
photon at $t$\,=\,0}
\par
Subtituting the amplitude in \eq{8} for the cross section, we have for
three flavours and three colours
 \beq \label{22}
x_P\,\frac{d \s}{ d x_P\,d t } \mid_{t = 0 }\,\,=\,\,\frac{4\,\pi^2}{3}\,
\,\alpha_{em}\,\as^2\sum_F
\,Z^2_F \,
\,\,\,\frac{1}{Q^4}\,\b^3\,
\,\,( 1 - 2 \b )^2
\eeq
$$
\,\,\int^{\frac{M^2}{4}}_{Q^2_0}\,\,\frac{ d\, k^2_{\perp}}{k^2_{\perp}}
\,\,
\(\,x_P\,G(\,\frac{k^2_{\perp}}{ \,(1 - \b)},x_P\,)\,\)^2
\frac{1}{\sqrt{1\,-\,\frac{4 k^2_{\perp}}
{M^2}}}
$$
\par
One can conclude from \eq{22} that only small distance
processes contribute
to the cross section. This fact manifests itself in the log
integration with respect to $k^2_{\perp}$. Even in the double log
approximation
of pQCD, this integral converges and can be rewritten in the form
\beq \label{23}
x_P\,\frac{d \s}{ d x_P\,d t } \mid_{t = 0 }\,\,=\,\,\frac{4 \pi^3}
{3 N_C}\,
\alpha_{em}\,\as^3 \,\sum_F Z^2_F
\,\,\frac{1}{Q^4}\,\b^3\,
\,\,( 1 - 2 \b )^2
\eeq
$$
\,\,\int^{1}_{x_P}\,\,\frac{ d\, x'_P}{x'_P}\,\,
\(\,\frac{ \partial (\, x_P \,G(\,\frac{M^2_{\perp}}{4\, \,(1 - \b)},x_P)
\,}
{\partial ln (1/x'_P)}\,\)^2
$$
The above formula gives the result for the case of a two jet production
originating from a $\bar q q $ pair.

\subsection{The DD amplitude at $q_{\perp}$ = 0
for a transverse photon}
\par
Using the formulae given in the previous subsections,
and taking into account eqs.(3)
and (4) for the gluon structure function, we obtain the following
expression for the amplitude $ f $:
\beq \label{59}
f\,\,=\,\,\int\,\, r_{\perp} d r_{\perp}\,\,\,
 a\,K_1(\,\frac{Q}{M} \,k_{\perp} \,r_{\perp}\,)
\,\,\frac{8 C_F \as }{N^2_C - 1}\,\,\pi^2
\int_C \frac{d \o}{ 2 \pi i} \,
\eeq
$$
\,g(\o)\,\,\g(\o) \,\,
(\frac{r^2_{\perp}}{4})^{1 - \g(\o)}\,\,\,\frac{\G(\g(\o)) \,\,
\G(1 - \g(\o))}{(\,\G(2 - \g(\o)\,)\,)^2}
\,\, J_1 (k_{\perp}\,\,r_{\perp})
$$
Making use of the  properties of the modified Bessel functions we
perform the integration over $r_{\perp}$ and obtain
\beq \label{60}
f\,\,=\,\,\,
 \frac{8C_F \as}{N^2_C - 1}\,\,\pi^2 \,\,k_{\perp}
\int_C \frac{d \o}{ 2 \pi i} \,\,g(\o)\,
\,\, \frac{1}{a^4}\,\,
(\frac{a^2}{\b})^{\g(\o)}\,\,
\eeq
$$
\,\,\frac{\G(1 + \g(\o)) \,\,
\G(1 - \g(\o)) \,\G(2 - \g(\o))\,\G(3 - \g(\o))}{\G( 2\, -\,g(\o))\,\G
(2)} \,\b^3 \,\,{}_1F_1 \(\, 3 \,-\,\g(\o), g(\o),
2, \,1\,-\,\b\,\)
$$
where $a \,=\,Q\,\frac{k_{\perp}}{M}$.
\par
In the DGLAP approach the above equation can be simplified as
$ \g(\o)\,\ll\,1$, it reduces to the form
\beq \label{61}
f\,\,=\,\pi^2 \,\frac{16\,C_F\,\as}{N^2_C - 1}\,\,\frac{k_{\perp}^2}
{a^4} \,\beta^3\,\,x_{P} \,G\(\,\frac{k^2_{\perp}}{( 1 - \b)},x_P\,\)
\eeq

\subsection{DD cross section for a transverse photon
into a $\bar q\, q$ pair}
Substituting the amplitude $f$ in the general formula for the cross
section (see \eq{8}), we obtain for three flavours
\beq \label{62}
x_P\,\frac{d \s}{ d x_P\,d t } \mid_{t = 0 }\,\,=\,\,
\frac{4\,\pi^2}{3}\,\,\alpha_{em}\,\as^2\,\,
\sum Z^2_F\,\,\frac{1}{Q^2}\,\b^3\,(\,1\,-\,\b)^2
\eeq
$$
\,\,\int^{\frac{M^2}{4}}_{Q^2_0}\,\,\frac{ d\, k^2_{\perp}}{k^4_{\perp}}
\,\,\(x_P\,G\(\,\frac{k^2_{\perp}}{\,(1 - \b)},x_P\)\,\)^2\,\,
\(\,1\, - \,2 \frac{k^2_{\perp}}{M^2}\,\)\,\,\frac{1}{\sqrt{1 -
 \frac{4k^2_{\perp}}{M^2}}}
 $$

\subsection{DD cross section for
{\bf $\bar q\, q\, G$} production}
The emission of one additional gluon is shown in the diagrams of Fig.2.
One can see two different classes of diagrams which
describe gluon emission. In the first, the emitted gluon
does not interact with the target as shown in Fig.2a. The general way
to take such gluon emission into account, is to use the evolution
equations.
We shall write down explicitly all formulae for the emission of one
gluon at  the end of this section.
The second class of diagrams
describes the process in which the emitted gluon interacts with the
target.
(see Fig.2b). These diagrams have to be calculated separately so as to
provide the initial conditions for the evolution
equations\cite{LW}.
\par
We start with the calculation of the diagrams of Fig.2b.
We would like to determine those contributions where the smallness
of $\as$ is compensated by a large logarithm. In other words, we are
looking for contributions of the order of $\as \ln ( Q^2/k^2)$. Our first
observation is that such a term does not exist in the case of a
longitudinal
polarised photon induced reaction. Indeed, this fact can  easily be
seen from the
general structure of the diagrams of Fig. 2b, namely, they can be written
in a general form
\beq \label{GE1}
x_P\,\frac{d \s}{ d x_P\,d t } \mid_{t = 0 }\,( Fig. 2b)\,\,\propto\,\,
\int^1_{0} \, d z \,\,\int^1_{x_P} \,\frac{d x'}{x'}
\int\,\frac{d^2 r_{\perp}}{2 \pi}\,\,
\Psi^{\g^*} \,\s(r_{\perp},q^2_{\perp} = 0,x' ) [\Psi^{\g^*}]^* \,\,
\eeq
where  $\s$ is defined by (see \eq{18})
\beq \label{GE2}
\s(r_{\perp},q^2_{\perp})\,\,=\,\,\frac{\as(\frac{4}{r^2_{\perp}})}{3}
\,\,\pi^2\,\,r^2_{\perp}\,\,
\(\,\, x G( \frac{4}{r^2_{\perp}},x )\,\,\)^2
\eeq
with
$[x G(r^2_{\perp},x)]^2$ that has been introduced in Ref.\cite{MUQI}.
The explicit expression for $(x G)^2$ was
derived in Ref.\cite{M90} and it is equal to
\beq \label{GE3}
\(x' G(r^2_{\perp},x')\)^2\,\,=\,\,\frac{2}{\pi^2}\,\,
\int^{\infty}_{r^2_{\perp}}
\,\,\frac{ d^2 r'}{r'^4}\,\int^{\infty}_ 0 \,\, d b^2_{\perp}
\,[\, \s^{GG}\,]^2
\,\,
\eeq
where
\beq \label{GE4}
\s^{GG}(r',x')\,\,=\,\,\frac{ 3\,\as}{4}\,\,r'^2\,\,\(\,x'\,
G(\frac{4}{r'^2},x')\,\)
\eeq
Substituting \eq{13} for $\Psi^{\g^*}_L$ in \eq{GE3},
after the integration over $z$ and using the properties of the modified
Bessel functions, we obtain
\beq \label{GE5}
\s(r_{\perp},q^2_{\perp})\,\,\propto \,\,\int^{\infty}_{\frac{4}{Q^2}}
 \,\,\frac{d r^2_{\perp}}{r^4_{\perp}}\frac{\as(\frac{4}{r^2_{\perp}})}
{3}
\,\,\pi^2\,\,\,
\(\,\, x G( \frac{4}{r^2_{\perp}},x )\,\,\)^2
\eeq
Therefore, we have no log contribution from the $r_{\perp}$ integration.
However, if we calculate the same integral for the transverse polarised
photon we find
\beq \label{GE6}
\s(r_{\perp},q^2_{\perp})\,\,\propto \,\,\int^{\infty}_{\frac{4}{Q^2}}
 \,\,\frac{d r^2_{\perp}}{r^2_{\perp}}\frac{\as}{3}
\,\,\pi^2\,\,\,
\(\,\, x G( \frac{4}{r^2_{\perp}},x )\,\,\)^2
\eeq
$$
\,\, \propto
\,\,\int^{\infty}_{\frac{1}{Q^2}} \,d r'^2 \,\ln (Q^2\,r'^2)\,\,
\(\,\, x G( \frac{4}{r^2_{\perp}},x )\,\,\)^2\,\,
$$
Finally, going to the $k_{\perp}$ representation, \eq{GE6} yields the
result of Ref.\cite{LW}
\beq \label{GE7}
x_P \,\frac{d \s_T}{dx_{P}d t}|_{t = 0}\,( Fig.2b)\,\,=\,\,
\sum_F \,\,\frac{ 4\,\pi^2\,\alpha_{em}\,Z^2_F}{Q^2}\,\,\b\,\int^{1}_{\b}
\,\frac{ d z}{z^4}\,\,
\int^{\frac{M^2}{4}}_{{\bar Q}^2_0}
\,\frac{d k^2}{k^4} \,\frac{\as^3 N^2_C}{ 32 \pi}
\eeq
$$
\,\,\{\,\b^2 \,+\,( z\,-\,\b )^2\,\}\,(1 - z )^3 \,(2 z + 1 )^2\,\,
\(\,x_P\,G( k^2,x_P )\,\)^2\,\,
$$
Taking the integral over $z$ we obtain
\beq  \label{GE8}
x_P \,\frac{d \s_T}{dx_{P}d t}|_{t = 0}\,( Fig.2b)\,\,=\,\,
\sum_F \,\,\frac{ 4\,\pi^2 \,\alpha_{em}\,Z^2_F}{Q^2}\,\,
\int^{\frac{M^2}{4}}_{{\bar Q}^2_0}
\,\frac{d k^2}{k^4} \,\frac{N^2_C \as^3 }{ 32 \pi}\,
\,\,\ln \frac{M^2}{4\,k^2}\,\,
\eeq
$$
\{\,\,\frac{(1 \,-\,\b)}{6}\,[\,4\,-\,31\,\b\,-\,63\,\b^2\,+\,50\,\b^3\,
-\,14\,\b^4\,]\,\,-\,\,\ln \b\,[\,1 \,+\,10\,\b\,-\,2\,\b^2\,]\,\}\,\,
\(\,x_P\,G(x_P, k^2 )\,\)^2
$$
\par
We now calculate the diagrams of Fig.2a. For a transverse polarized
photon, this has been done in Ref. \cite{LW} and the result in our
notation is
\beq \label{GE9}
x_P \,\frac{d \s_T}{dx_{P}d t}|_{t = 0}\,( Fig.2a)\,\,=\,\,
\sum_F \,\,\frac{ 4\,\pi^2 \,\alpha_{em}\,Z^2_F}{Q^2}\,\,\beta\,\int^{1}_
{ \b}
\,\frac{ d z}{z^5}\,\,
\int^{\frac{M^2}{4}}_{{\bar Q}^2_0}
\,\frac{d k^2}{k^4 } \,\frac{4\,\as^3 }{3 N_C \pi}\,
\,\ln\frac{M^2}{4\,k^2}\,\,
\eeq
$$
\,\frac{1 + z^2}{1 - z}\,\,\{  \b^2\,( z - \b)^2
\,\,-\,\,z^6 \,\b^2 (1 - \b)^2\,\}
\,\,\(\,x_P\,G( k^2,x_P )\,\)^2\,\,\,\,=
$$
$$
\sum_F \,\,\frac{ 4\,\pi^2 \,\alpha_{em}\,Z^2_F}{Q^2}\,\,
\int^1_\b \,\,
\int^{\frac{M^2}{4}}_{{\bar Q}^2_0}
\,\frac{d k^2}{k^4 } \,\frac{4\,\as^3 }{3 N_C \pi}\,
\,\ln\frac{M^2}{4\,k^2}\,\,
$$
$$
\b\,( 1\,-\,\b)^2\,\{\,-\,2\,\b \,\ln \b \,\,+\,\,\frac{1}{12}\,[\,1\,+\,6\,\b
\,-\,9\,\b^2\,-\,24\,\b^2\,-\,6\,\b^4\,-\,4\,\b^5\,]\}\,\,
\(\,x_P\,G( k^2,x_P )\,\)^2\,\,
$$
\par
One can  calculate
the one gluon emission, for the longitudinal polarised photon,
 using the general formula obtained in Ref.\cite{LW}
\beq \label{GE10}
x_P \,\frac{d \s_{L}}{dx_{P}d t}|_{t = 0}\,( Fig.2a)\,\,=\,\, \sum_F
 \,\,\frac{ 4\,\pi^2 \,\alpha_{em}\,Z^2_F}{Q^2}\,\,\b\,\int^{1}_{\b}
\,\frac{ d z}{z}\,\,
\eeq
$$
\int^{\frac{M^2}{4}}_{{\bar Q}^2_0}
\,\frac{d k^2}{k^4} \,\frac{\as^3 }{ 32 \pi}\,
\,\ln \frac{Q^2}{k^2}\,\,\Phi^F_F ( \frac{\b}{z} )\,\,\Phi^F_{P; T} ( z)
\(\,x_P\,G( k^2,x_P )\,\)^2\,\,
$$
where $\Phi^F_F$ denotes the usual DGLAP splitting function,
and $\Phi^F_P$ is the
splitting function of the Pomeron into a quark-antiquark pair. For
the transverse
polarisation of the photon this splitting function has been calculated
previously in Refs.\cite
{NZ1}\cite{LW} and we have recalculated it in the $r_{\perp}$ -
representation (see \eq{62})
\beq \label{PHIT}
\Phi^F_{P;T}\,\,=\,\,\frac{16}{N_C}\,z^2\,(1 - z)^2\,\,
\eeq
For the longitudinal polarization we
obtain the same factorized answer with one important difference,
the integration over $k^2$ is logarithmic (see \eq{22}).
However, as far as the z - dependence is  concerned, the formula is the
same as
\eq{GE10} with the splitting function which we have calculated
previously (see \eq{20})
and which is equal to
\beq \label{PHIL}
\Phi^F_{P;L}\,\,=\,\,\frac{16}{N_C}\,z^3\,(1 - 2 z)^2\,\,
\eeq
Finally, the answer for the one gluon emission in the case of
a longitudinal polarised photon is
\beq \label{GE11}
x_P \,\frac{d \s_L}{dx_{P}d t}|_{t = 0}\,( Fig.2a)\,\,=\,\,
\sum_F \,\,\frac{ 4\,\pi^2 \,\alpha_{em}\,Z^2_F}{Q^4}\,\,
\int^{1}_{\b} \,\frac{ d z}{z^6}\,\,
\int^{\frac{M^2}{4}}_{{\bar Q}^2_0}
\,\frac{d k^2}{k^2} \,\frac{4\,\as^3 }{3\, N_C \pi}\,
\,\ln\frac{Q^2}{k^2}\,\,
\eeq
$$
\b\,\,\,\frac{1 + z^2}{1 - z}\,\,\{  \b^3\,( z - 2 \b)^2
\,\,-\,\,z^7 \,\b^3\, (1 - 2 \b)^2\,\}
\(\,x_P\,G( k^2,x_P )\,\)^2\,\,=
$$
$$
\sum_F \,\,\frac{ 4\,\pi^2 \,\alpha_{em}\,Z^2_F}{Q^4}
\,\int^{\frac{M^2}{4}}_{{\bar Q}^2_0}
\,\frac{d k^2}{k^2} \,\frac{4\,\as^3 }{3\, N_C \pi}\,
\,\ln\frac{Q^2}{k^2}\,\,
$$
$$
\{-\,2\,\b^3\,\ln \b\,( 1 - 2 \b)^2\,+\,\frac{2}{15}\,+\,\frac{1}
{6} \,\b
\,+\,\frac{2}{3}\,\b^2\,-\,4\,\b^3\,+\,9\,\b^4\,+\,\frac{74}{15}\,\b^5\,
-\,\frac{19}{3}\,\b^6\,-\,\frac{2}{3}\,\b^7\,\}
\,\(x_P\,G(k^2,x_P)\)^2
$$

\section{Shadowing corrections for the diffractive production of small
masses}
\par
The advantage of using
the $r_{\perp}$ representation is not apparent in
the calculations performed in the previous section, in that we
obtain the same results previously
obtained by using
momentum space calculation techniques (see for example Refs.\cite
{NZ1}\cite{RYS}\cite{LW}\cite{GLR}\cite{MUQI}).
The above calculations are instructive and
they provide an explicit example of the relations between
different variables, used to describe deep
inelastic process in different frames. In this section,
where we calculate the shadowing corrections for the various processes,
the
$r_{\perp}$-representation simplifies the calculation.
We will only calculate the damping factors for the penetration of a
$\bar q q $ - pair through the target, assuming that all SC
to the gluon density have been included in the
phenomenological gluon density that we use
in the
calculations. The size of the corrections due to damping in the gluon
sector has been estimated in Ref.\cite{GLMVEC}.

\subsection{The $b_{\perp}$ dependance of the amplitude}
\par
To estimate the  shadowing corrections, we need to know the profile
of the amplitude in impact parameter space, which means that we need to
know the amplitude not only at
$q_{\perp}$=0, but also at all values of the momentum transfer.
The gluon density is weakly dependent on $q_{\perp}$
at small values of $q_{\perp}$,
both in the double log approximation (see a detailed discussion in
Ref.\cite{GLR}) and in the BFKL approach (see Refs.\cite{LRREV}\cite
{LRSLOPE}). Therefore, the leading $q_{\perp}$-dependence comes
from
the form factor of the $\bar q q$ pair with a transverse separation of
$r_{\perp}$, and the form factor of the target (proton).
\par
As the form factor cannot be
treated theoretically in pQCD, we will assume an
exponential parametrization for it, namely
\beq \label{Fp}
F_p ( q^2_{\perp})\,\,=\,\,e^{- \frac{B}{4} \,\,q^2_{\perp}}
\eeq
The slope $B$ can be extracted from the experimental data on
hadron-hadron
collisions, provided we take  the Pomeron slope $\a'$=0. Namely,
 $B$ = $B^{pp}_{el}(\a' = 0 )$, where $B_{el}$ is the slope in the
differential
cross section of the proton-proton collision\cite{GLM}.
It turns out that the value of $B$ deduced from the hadronic data is very
close to
the one obtained from the proton electromagnetic form factor.
\par
The form factor of the $\bar q q $ pair with a transverse separation
of $r_{\perp}$ is equal to
\beq \label{40}
F_{\bar q q} ( q^2_{\perp})\,\,=\,\,\Psi^{i}_{\bar q q} \(
\frac{({\vec{k}}_{1\perp}\,\,-\,\,{\vec{k}}_{2\perp})\,\cdot\,
{\vec{r}}_{\perp}}{2} \) \,\,\,\,\Psi^{f*}_{\bar q q} \(
\frac{({\vec{k'}}_{1\perp}\,\,-\,\,{\vec{k'}}_{2\perp})\,
\cdot\,{\vec{r}}_{\perp}}{2} \)
\eeq
where $k_i$ ($k'_i$) denote the momentum of the quark $i$ before
and after the collision. Each of the wave functions is given by
the exponent and a
simple sum of different attachments of gluon lines to quark lines.
Hence,
\beq \label{41}
F_{\bar q q} ( q^2_{\perp})\,\,=\,\,e^{ i \frac{ {\vec{q}}_{\perp}\,\cdot
\,{\vec{r}}_{\perp}}{2}}\,\,\,\,\(\,\,1\,\,-\,\,e^{i\,\vec{l}_{\perp}
\cdot\vec{r}_{\perp}}\,\,\)
\eeq
The last factor is absorbed in the expression for the cross section,
while the first factor gives the $q_{\perp}$ dependence of the $\bar q q$
form factor, which after integration over the azimuthal angle is
\beq \label{42}
F_{\bar q q} ( q^2_{\perp})\,\,=\,\,J_0(\frac{q_{\perp}\,\,\,r_{\perp}}
{2})
\eeq
To proceed with the calculation we require the profile function
in $b_{\perp}$ space
\beq \label{43}
S(b^2_{\perp})\,\,=\,\,\frac{1}{4\pi^2} \,\,\int \,\,d^2  q_{\perp}\,\,
e^{i {\vec{b}}_{\perp}\,\cdot\,{\vec{q}}_{\perp}}\,\,\,F_p ( q^2_{\perp})
\,\, F_{\bar q q} ( q^2_{\perp})
\eeq
The explicit calculation gives
\beq \label{44}
S(b^2_{\perp})\,\,=\,\,\frac{1}{\pi B}\,\,I_0 \(\frac{b_{\perp}\,\,r_
{\perp}}{B}
\)\,\,e^{-\,\,\frac{b^2_{\perp}\,\,+\,\,\frac{r^2_{\perp}}{4}}{B}}
\eeq
To simplify the calculation we replace the above function by
\beq \label{45}
S(b^2_{\perp})\,\,=\,\,\frac{1}{\pi B'}\,\,e^{- \frac{b^2_{\perp}}{B'}}
\eeq
where
\beq \label{BP}
B'\,\,=\,\,B\,\(\,1\,\,+\,\,\frac{r^2_{\perp}}{4 B}\,\)\,\,\approx\,\,
B\,\(\,1\,\,+\,\,\frac{1}{a^2\, B}\,\)
\eeq
For sufficiently large values of $a$, we
consider $a^2 B \,\,\gg \,1$, and neglect the second term in our
calculation.

\subsection{Penetration of a $\bar q q $-pair through the target}
\par
To calculate the shadowing correction we follow the procedure suggested
in Refs.\cite{LR}\cite{M90}.
Namely, we replace $\s(r_{\perp},q^2_{\perp} = 0 )$ in \eq{12} by
\beq \label{46}
\s^{SC}(r_{\perp})\,\,=\,\,2\,\,\int d^2 b_{\perp} \,\,\( \,\,1\,\,-\,\,
e^{-\,\frac{1}{2}\,\,\s(r_{\perp},q^2_{\perp} = 0 )\,\,S(b^2_{\perp})}
\,\,\)
\eeq
The above formula is the solution of the s-channel unitarity constraint
\beq \label{47}
2\,\,Im\,a(s,b_{\perp})\,\,=\,\,|a(s,b_{\perp})|^2 \,\,+\,\,G_{in}(s,b_
{\perp})
\eeq
where $a$ denotes the  elastic amlitude for $\bar q q $ pair with a
transverse
separation $r_{\perp}$, while $G_{in}$ is the contribution of the
inelastic processes. The inelastic cross section is equal to
\beq \label{48}
\s_{in}\,\,=\,\,\int \,\,d^2 \,b_{\perp}\,\,G_{in} ( s, b_{\perp})
\,\,=\,\,\int d^2 b_{\perp} \,\,\( \,\,1\,\,-\,\,
e^{-\,\,\s(r_{\perp},q^2_{\perp} = 0 )\,\,S(b^2_{\perp})}\,\,\)
\eeq
Our formulation
is based on the physical assumption that the
structure of the final state is obtained from
the uniform parton distribution derived
from the QCD evolution equation. We neglect
the contribution to the inelastic final state of all DD
processes with
large rapidity gap.
For example, "fan" diagrams
that give an important contribution are neglected\cite{GLR}
as well as DD in the region of small mass, which
cannot be presented as the decomposition of the $\bar q q$ wave function.
\par
>From a point of view of Feynman diagrams, \eq{48} sums all the
diagrams of Figs. 1 and 2  in which the $\bar q q$ -pair rescatters off
the target and exchanges
many "ladder" diagrams, each of which
can be represented by the gluon density.
This sum has been performed by Mueller\cite{M90}, and we shall
perform the calculation for the case of small mass DD.

\subsection{Damping factor for diffractive production of small masses
induced by a longitudinal photon.}
Substituting $\s^{SC}$ of \eq{46} in
\eq{12} and using
$\s$ in the form of \eq{18},
enables us to
estimate the general term of the expansion with respect to $\s$.
It has the form
\beq \label{49}
f^{n}\,\, = \,\,C \,\, \frac{k^2_{\perp}}{2Q}\,\,\,\,
\int \,\frac{d^2 \,r_{\perp}}{\pi} \,\,K_0 ( a \,r_{\perp})
\frac{( - 1 )^{n- 1}}{n!}
\,\,\(\frac{\as\,4 C_F \pi^2}{N^2_C - 1}\)^n \,\,
\eeq
$$
\prod^n_{i}\,\,\int_{C_i} \,\,\frac{d\,\o_i\,e^{\sum \o_i \ln(1/x_P)}
}{2 \pi i}\,\,g_i (\o_i)\,\,
\,\frac{\G(1 + \g(\o_i))\,\,\G(1 - \g(\o_i))}{(\G(2 - \g(\o_i)))^2}
$$
$$
\,\,\,\,\(\frac{r^2_{\perp}}{4}\)^{n \,\,-\,\,\sum^n_i \,\g(\o_i)}
\,\,\,J_0 ( k_{\perp} \,\,r_{\perp})
 \,\, \int \,d^2 b_{\perp}\,\, S^n(b^2_{\perp})
$$
We replaced all numerical coefficients by the factor $C$.
\par
Integrating over $r_{\perp}$ and $b_{\perp}$ we obtain the
expression
\beq \label{50}
f^{n}\,\, =\,\, C  \frac{1}{2 Q}\,\,\,\,
\frac{( - 1 )^{n - 1}}{n\,\,n!}
\,\,\(\frac{\as\,4\, C_F \pi}{B'(N^2_C - 1)}\)^n
\eeq
$$
\prod^n_{i}\,\,\int_{C_i} \,\,\frac{d\,\o_i\,e^{\sum \o_iln(1/x_P)}}{2
\pi i}
\,\,g_i (\o_i)\,\,
\,\,\frac{\G(1 + \g(\o_i))\,\,\G(1 - \g(\o_i))}{( \G(2 - \g(\o_i)))^2}
$$
$$
\,\,( \frac{a^2}{\b})^{\sum \g(\o_i)}\
\,\,\G^2( 1 + n - \sum^n_i \g(\o_i))\,\,\(\,\frac{\b^2}{a^2}\,\)^n
\,\,{}_1F_1\(1 + n -\sum \g(\o_i),- n +\sum \g(\o_i),1, \,1\,-\,\b\,\)
$$
\par
In the double log approximation of pQCD with $\g(\o_i)\,\,\ll\,\,1$,
neglecting $\g$ and taking
$ 1 - \b\,\ll \,1$, we derive a very simple formula.
Taking the integral over $\o_i$
yields
\beq \label{51}
f^{n} = C \,\,B'\,\, \frac{\b^2}{2\,\,Q}
\,\,
\frac{( - 1 )^{n - 1}n!}{n}
\,\,\(\frac{ C_F \pi}{B'(N^2_C - 1)}\,\,\frac{4 \b}{a^2} \,\,\as\,\,
x_P G(\frac{a^2}{\b},x_P) \)^n
\eeq
Finally, for $f$ we have
\beq \label{52}
f\,\,=\,\,C\,\, B'\,\,\sum^{\infty}_{n = 1}
 \frac{\b^2}{2\,Q}\,\,(n - 1 )!\,\,
\(\frac{ C_F \pi}{B'(N^2_C - 1)}\,\,\frac{4\b}{a^2} \,\,\as\,\,x_P
 G(\frac{a^2}{\b},x_P) \)^n
\eeq
The above series can be summed to give the analytic function $E_1$,
namely
\beq \label{53}
f\,\,=\,\,C\,\, B'\,\,
 \frac{\b^2}{2\,\,Q}\,\,E_1 (\,\frac{1}{\kappa_q}\,)\,\,
e^{\frac{1}{\kappa_q}}
\eeq
where (for $N_C$= 3)
\beq \label{54}
\kappa_q\,\,=\,\,\frac{2}{3}\,\,\frac{\as \pi\,\b}{B' \,\,a^2} \,\,
\,\,
x_P G(\frac{a^2}{\b}, x_P)\,\,=\,\,\frac{2}{3}\,\,
\frac{\as \pi\,( 1 - \b)}
{B' \,\,k^2_{\perp}} \,\,
\,\,
x_P G\(\frac{k^2_{\perp}}{(1 - \b)},x_P\)
\eeq
Using the above equation, we obtain the following expression for the
damping factor
\beq \label{55}
D^2_L\,\,=\,\,\frac{\int^{\frac{M^2}{4}}_{Q^2_0}\,\,k^2_{\perp}\,d\,k^2_
{\perp}\,\,E^2_1 (\,\frac{1}{\kappa_q}\,)\,\,
e^{\frac{2}{\kappa_q}}}{\int^{\frac{M^2}{4}}_{Q^2_0}\,
\,k^2_{\perp}\,d\,k^2_{\perp}
\,\,\kappa^2_q\,}
\eeq
The behaviour of the damping factor for small and large $\kappa_{q}$
can be
found by using the well known
property of $E_1$. Namely, for
$\kappa_q \,\ll\,1$, $D \ra 1$, while at $\kappa_q\,\gg\,1$ the damping
factor vanishes as
\beq \label{DL2}
D^2_L\, \propto \,\frac{\int^{\frac{M^2}{4}}_{Q^2_0} \,k^2_{\perp}
 d\,k^2_{\perp}
\ln^2 \kappa_q}{\int^{\frac{M^2}{4}}_{Q^2_0} \,k^2_{\perp}
 d\,k^2_{\perp}\kappa^2_q}\,
\eeq
The behaviour of $D_L^2$
as a function of $\b$  and $Q^2$ is given in Fig.3.
One can see that the value of the damping factor reaches 0.5 in the
region
of small $\b$ and $Q^2$. This means that the SC should be included even
for a longitudinal photon, where the smallest distances of the
order of $\frac{1}{Q}$ contributes.

\subsection{Damping factor for diffractive production of small masses
induced by a transverse photon}
\par
Repeating the steps of section 3.3
we derive the formula for the damping factor for a transverse
polarized photon. Substituting
$\s^{SC}$ of \eq{46} in \eq{12} we obtain
\beq \label{63}
f^{n}\,\, = \,\,C \,\, a\,\,\,\,
\int \,\frac{d^2 \,r_{\perp}}{\pi} \,\,K_1 ( a \,r_{\perp})
\frac{( - 1 )^{n- 1}}{n!}
\,\,\(\frac{\as\,4 C_F \pi^2}{N^2_C - 1}\)^n \,\,
\eeq
$$
\prod^n_{i}\,\,\int_{C_i} \,\,\frac{d\,\o_i\,e^{\sum \o_i \ln(1/x_P)}
}{2 \pi i}\,\,g_i (\o_i)\,\,
\,\,\frac{\G(1 + \g(\o_i))\,\,\G(1 - \g(\o_i))}{( \G(2 - \g(\o_i)))^2}
$$
$$
\,\,\,\,\(\frac{r^2_{\perp}}{4}\)^{n \,\,-\,\,\sum^n_i \,\g(\o_i)}
\,\,\,J_1 ( k_{\perp} \,\,r_{\perp})
\,\, \int \,d^2 b_{\perp}\,\, S^n(b^2_{\perp})
$$
$C$ stands for all numerical coefficients.
\par
After integration over $r_{\perp}$ and $b_{\perp}$ we have
\beq \label{64}
f^{n}\,\, =\,\, C  \,\b \frac{k_{\perp}}{a^2}\,\,\,\,
\frac{( - 1 )^{n - 1}}{n\,\,n!}
\,\,\(\frac{\as\,4\, C_F \pi}{B'(N^2_C - 1)}\)^n
\eeq
$$
\prod^n_{i}\,\,\int_{C_i} \,\,\frac{d\,\o_i\,e^{\sum \o_iln(1/x_P)}}
{2 \pi}
\,\,g_i (\o_i)\,\,
\,\,\frac{\G(1 + \g(\o_i))\,\,\G(1 - \g(\o_i))}{( \G(2 - \g(\o_i)))^2}
\,\,( \frac{a^2}{\b^2})^{\sum \g(\o_i)}\,\(\,\frac{\b^2}{a}\,\)^n
$$
$$
\,\,\G( 2 + n - \sum^n_i \g(\o_i))\,\,\G(1 + n - \sum^n_i \g(\o_i))\,\,
{}_1F_1\( 2 + n - \sum \g(\o_i), 1 - n +\sum \g(\o_i),2, \,1\,-\,\b\,\)
$$
Again,
in the double log approximation of pQCD with $\g(\o_i)\,\,\ll\,\,1$,
neglecting $\g$ as well as considering
$ 1 - \b\,\ll\,1$, we take the integral over $\o_i$
and get
\beq \label{65}
f^{n} = C \,\,B'\,\, \frac{\b k_{\perp}}{a^2}
\,\,\,\,
\frac{( - 1 )^{n - 1}n!(n + 1)}{n}
\,\,\(\frac{ C_F \pi}{B'(N^2_C - 1)}\,\,\frac{4 \b}{a^2} \,\,\as\,\,
x_P G(\frac{a^2}{\b},x_P) \)^n
\eeq
Finally we obtain
\beq \label{66}
f\,\,=\,\,C\,\, B'\,\,\sum^{\infty}_{n = 1}
 \frac{\b k_{\perp}}{ a^2}\,\,\(\,(n - 1 )! \,+\,n!\,\)
\(\frac{ C_F \pi}{B'(N^2_C - 1)}\,\,\frac{4\b^2}{a^2} \,\,\as\,\,x_P
 G(\frac{a^2}{\b^2},x_P) \)^n
\eeq
The above series can written in terms of the analytic function $E_1$
\beq \label{67}
f\,\,=\,\,C\,\, B'\,\,
 \frac{\b k_{\perp}}{ 2\,a^2}\,\,\(\, 1\,\,+\,\,
(\,1 \,-\,\frac{1}{\kappa_q} \,)\,\,E_1 (\,\frac{1}{\kappa_q}\,)\,\,
e^{\frac{1}{\kappa_q}}\,\)
\eeq
where $\kappa_q$ is given by \eq{54}.
\par
Using the above equation we get
\beq \label{68}
D^2_T\,\,=\,\,\frac{\int^{\frac{M^2}{4}}_{Q^2_0}\,\,d\,k^2_{\perp}
\{\, 1\,\,+\,\,
\(\,1 \,-\,\frac{1}{\kappa_q} \,\)\,\,E_1 (\,\frac{1}{\kappa_q}\,)\,\,
e^{\frac{1}{\kappa_q}}\,\}^2}{4\,\,\int^{\frac{M^2}{4}}_{Q^2_0}\,
\,\,d\,k^2_{\perp}
\,\,\kappa^2_q\,}\,\,
\eeq
Fig. 4 shows the dependence of $D^2_T$ on $\b$ and $Q^2$.
It is interesting
to note that the value of $D^2_T$ is small for small $\b$. This
reflects a well
known fact that the SC are large for DD producing a
system with a large mass \cite{GLMD}\cite{GLR}.

\subsection{The relationship  between SC, $F_2 (Q^2, x_B)$ and
the DD processes}
The simplest relation between the corrections to $F_2$ and
the DD cross section
can be derived directly from the AGK cutting rules\cite{AGK}
(see ref.\cite{LW}) and it reads
\beq \label{86}
\frac{\Delta F_2 (Q^2, x_B)}{F_2 (Q^2, x_B)}\,\,=\,\,\frac{\s^{DD}}{\s_
{tot}}
\eeq
where
\beq \label{87}
F_2\,\,=\,\,F^{DGLAP}_{2} \,\,-\,\,\Delta F_2\,\,
\eeq
However, \eq{86} is only valid when the DD cross
section is small.
The notion  of what is meant by "small" in diffractive production
is not unique, since DD is small in two cases:
i) when the kinematic region between partons is very
small, this  can be dealt with in a perturbative way, or
ii) in the  case when the interaction is so strong that we have
scattering off a black disc.
As we have shown, the damping factors for DD turn out to be rather
large. It is therefore necessary to reconsider the simple relation in
\eq{86}.
\par
We generalise \eq{86} by calculating the SC for $F_2$ due to the
penetration of a
$\bar q q $ pair through the nucleon. As we have mentioned above, we
adopt throughout the paper
the approach that the SC in the gluon sector have been already
taken into account in the phenomenological set of densities that
we use\cite{GRV}. The expression for the $F_2$ including
SC for $\bar q q $ pair has been derived by Mueller\cite{M90} and
in our notation has the following form
\beq \label{88}
F_2(Q^2,x)\,\,=\,\,\sum_F Z^2_F \,\int^1_0 \,d\,z \int \,
\frac{d^2 r_{\perp}}{2 \pi}\,\,\Psi^{\g^*}_{T}( Q^2, r_{\perp}, z )\,\,
\s^{SC}( r_{\perp}) \,\,[ \Psi^{\g^*}_{T}( Q^2, r_{\perp}, z )]^*\,\,
\eeq
As was shown in ref.\cite{M90}, whithin the LLA of pQCD which we use
throughout this paper, we can safely replace $\Psi_{T} \Psi^*_{T}$ by
$\frac{1}{r^4_{\perp}}$ after integrating over $z$ in \eq{88}.
Mueller's formula finally reads (for $N_C$ = $N_F $ = 3)
\beq \label{89}
F_2(Q^2,x)\,\,=\,\,\sum_F Z^2_F \,\frac{6}{\pi^2} \int\,
\frac{d^2 b_{\perp}}{\pi}\,\,\int^{\infty}_{\frac{4}{Q^2}}\,\frac{d^2 r_
{\perp}}
{\pi} \,\frac{1}{r^4_{\perp}} \,\(\,\,1\,\,-\,\,
\exp(\, - \frac{1}{2}\,
   \s(r_{\perp},q^2_{\perp} = 0 )\,S(b^2_{\perp})\,)\)
\eeq
\par
Adopting the same procedure of integration as for \eq{55} and \eq{68},
we derive the formula for the elastic damping factor which we define by
\beq \label{90}
D^2_{el}\,\,=\,\,\frac{| \Delta F_2 | }{| F^{NBA}_2 |}\,\,
\eeq
where $F^{NBA}_2$ denotes the correction to $F_2$ due to two Pomeron
exchange. In
other words, $F^{NBA}_2$ is the next order to the Born approximation of
\eq{89}. The formula for $D^2_{el}$ is
\beq \label{91}
D^2_{el}\,\,=\,\,\frac{|\,(\,1 + \kappa_q\,)\,(\,E_1(\kappa_q)\,+\,
\ln \kappa_q
\,+\,C\,)\,\,+\,\,1\,\,-\,\,e^{ - \,\kappa_q}\,\,-\,\,2\,\kappa_q\,|}
{\frac{\kappa^2_q}{4}}\,\,
\eeq
where $\kappa_q$ is defined by \eq{54}
at $\frac{a^2}{\b}\,=\,Q^2$. $D^2_{el}$ is plotted in Fig.5
for different values of $Q^2$.
It is interesting to note that $D^2_{el}$ turns out to be very
close to 1 over a wide range of $x_P$
because $D^2_{el}\,\,=\,\,1 - \frac{\kappa_q}{9}$ at small values
of $\kappa_q$.
\par
Using $D^2_{el}$ and $D^2_T (D^2_L)$ we can rewrite \eq{86} in a more
general form
\beq \label{92}
\frac{\Delta F_2 (Q^2, x)}{F_2 (Q^2,x)}\,\,=\,\,
\eeq
$$
\,\frac{1}
{\s_{tot}(Q^2,x)}\,\,\int^1_0 d \b \frac{d\,\s^{DD}_T(Q^2,x, \b)}
{d x_P}\,\,\frac{D^2_{el}}{D^2_T}\,\,+\,\,\frac{1}{\s_{tot}(Q^2,x)}\,
\,\int^1_0\,d\,\b \,\frac{d \s^{DD}_L(Q^2,x,\b)}{d x_P}\,
\frac{D^2_{el}}{D^2_L}
\,\,
$$
>From Figs.3 and 4 one can see that the ratio $\frac{D^2_{el}}{D^2_T}$
can be rather large, and reaches a value of about 2 at low $x$
in the HERA kinematic region.

\section{Numerics and results}
\subsection{What distances are essential in DD processes?}
\par
The first question that we wish to address is, "what
are the distances which  are important in
DD processes?"
\newline
1. In the case of a longitudinal polarised photon, the answer
is obvious, just by examining \eq{22}. Indeed, in all calculations
that lead to \eq{21} and \eq{22} the typical distances are $r_{\perp} \,
\propto
\,\frac{1}{a} \,=\,\frac{M}{Q \,k_{\perp}}$. Since the integral over
$k_{\perp}$
in \eq{22} is logarithmic, we conclude that $k_{\perp} \,\approx \,M/2$,
from which
we estimate that the dominant distances are $r_{\perp}\,
\approx \,\frac{2}{Q}$.
Hence, we can safely utilize the pQCD approach
for inclusive production initiated by a
longitudinal
polarised photon. In Fig.6 we
plot our predictions for $\b $ = 0.8. In spite of the $1/Q^2$ suppression
we  obtain a large value for the cross section. This suggests an
alternate
way to extract the gluon density, similar to that suggested for
vector meson production\cite{FIVE}\cite{RYS}\cite{GLMVEC}. It has
several advantages:
(i) The cross section is larger than that for vector meson production;
(ii) The prediction is independent of the form chosen for the
nonperturbative wave function of the
produced vector mesons, which is an inherent
difficultly when making a definite prediction
for the DD cross section for vector meson production \cite{GLMVEC}\cite
{FKS}; (iii) The seperation of the longitudinal from the tranverse
polarised photon is not difficult. We have not found any
contamination fron a transverse polarised photon induced reaction,
for events with large values of $\b$ (practically for $ \b \,>\,0.7$).
\newline
2. In the case of a transverse polarised photon, the value of the
distances for which the pQCD
calculation of the DD cross section is valid,
depends crucially on the behaviour of the gluon density function
at relatively small values of the photon virtuality. Indeed, the cross
section of \eq{62} has an extra $k^2$ in the denominator,
and the integral on first sight appears to be
infrared divergent. This is not so, as the gluon density
is proportional to $k^2/\mu^2$ at small values of $k^2$.
This fact is a direct consequence of the gauge invariance of QCD.
However, there is a danger that at very small values of $k^2$
in the nonperturbative QCD region, we could still have divergencies as
$\mu^2$ is about the size of the confiniment radius.
We know that $\mu$ depends on $x$, and
tends to be large at very small $x$ \cite{M90}\cite{GLR}. The question
arises as to what is
the situation in the kinematic region of HERA?
In an attempt to answer this
question we adopt the following strategy: We believe that the available
parametrizations of the DIS structure functions (GRV\cite{GRV},
MRS\cite{MRS}, CTEQ \cite{CTEQ}) describe the behaviour of the
gluon density in the region of relatively small virtualities.
The common feature of all these parametrization is the fact that
the behaviour $x G(k^2,x) \,\propto \,k^2$ starts at virtualities which
are sufficiently large in the small $x$ region. For larger $x$ and at
larger values of
$k^2$, $x G(k^2,x)$ starts to be proportional
to $k$ ( $x G(x,k^2) \,\propto\,
k/\mu $). For example, at $Q^2 = 2.5 GeV^2$, in the GRV parametrization,
$x G(x,k^2) \,\propto\,
k/\mu $ at $x \,<3. 10^{-3} $, while in the MRS(A) this happens at
$x\,< 10^{-2}$. Relying on  the available sets of parton densities
(which contain all accumulated experimental
information on the subject), we conclude that even for a transverse
polarized photon, the typical distances in the DD processes are not
larger than $r_{\perp} \,\approx \,1 GeV^{-1}$. Hence,
we can apply the
pQCD approach in this case as well.
\par
To illustrate this we plot in Fig.7 the integrand of \eq{62}
\beq \label{I}
I(k^2,x_P,\b)\,\,=\,\,\frac{ \left(\, x_P G(\frac{k^2_{\perp}}{1-\b},x_P
)\,\right)^2}{k^2_{\perp}}\,\,\(\, 1\,-\,2\,\frac{k^2_{\perp}}{M^2}\,\)\,
\,\frac{1}{\sqrt{1\,-\,\frac{4\,k^2_{\perp}}{M^2}}}\,\,
\eeq
as
a function of $x_P$ and $k^2$ at fixed $\b$.
We wish to emphasise that the integrand can be viewed
as the product of two factors:
$\(x_P G(\frac{k^2_{\perp}}{1 - \b},x_P)\)^2/k^2_{\perp}$
and the  kinematic factor $ (\,1 - 2\,\frac{k^2_{\perp}}{M^2}\,)\,\,
\frac{1}{\sqrt{1 - \frac{4 k^2_{\perp}}{M^2}}}$. The first factor has
a maximum at $k^2_{\perp} \,\approx \,1 \,GeV^2 $ for all available
parametrizations of the gluon density. The origin of this
maximum is very simple and can easily be traced if one uses the
semiclasical
parametrisation of the solution to the evolution equations in the form
\beq \label{SEMI}
x_P\,G( k^2,x_P )\,\,\rightarrow\,\,(\frac{1}{x_P})^{\o(k^2,x_P)}\,\,
(k^2)^{\gamma(k^2,x_P)}\,\,\,\,\,at\,\,\,\,x_P\,\,\rightarrow\,\,0\,\,
\eeq
where both $\o$ and $\g$ are smooth functions of $\ln(1/x_P)$ and
$\ln k^2$.
The common feature of the various parametrisations of the gluon density
is the fact that
$\g \,\,>\,\,\frac{1}{2}$ at $x_P \,\,\leq\,\,10^{-3}$.
Such a behaviour manifests itself in a maximum of the first factor at
$k^2\,
\approx\,1\,GeV^2$. It should be stressed that the argument of the gluon
density in \eq{62}, namely $k^2_{\perp}/( 1 - \b)$, leads to a
substantial increase of the typical transverse momentum in the integral.
This amounts to a decrease of the typical value of the distances
important
for diffractive $\bar q\, q $ production. The kinematic factor also
tends to increase at large values of $k^2$.
\par
Finally, we would like to repeat our statement that the DD process
initiated by
either longitudinal or transverse polarised photons occur
at small
distances. This fact is essential to justify the
use of pQCD for the calculation of DD.

\subsection{The $x_P$ dependence, factorization and the "Pomeron
structure function"}
\par
The $x_P$ dependence of the calculated $F^{DD(3)}_2$ is plotted
in Fig.8 together with the H1 data\cite{H1DDNEW}, where
\beq \label{DD(3)}
x_P\,F^{DD(3)}_2 ( Q^2,x_P,\b)
\,\,=\,\,\frac{Q^2}{4 \,\pi^2 \,\,\alpha_{em}}\,\,\int \,
d t \,\,x_P\,\frac{d \s}{d x_P d t }\,\,
=\,\,\frac{Q^2}{4 \,B\,\pi^2 \,\,\alpha_{em}}
\,\,x_P\,
\frac{d^2 \s}{d x_P d t }|_{t = 0}\,\,
\eeq
and $B$ denotes the slope of the DD cross section. We take
$B = 4.5\,GeV^{-2}$ in accord with the the prelimenary experimental data
from HERA
(see Ref.\cite{SLOPE}). The value of $B$ has not yet
been measured with good accuracy. The above  value of $B$ coincide with
the slope that has been extracted from the high energy phenomenology of
"soft" processes (see, for example, Ref.\cite{GLM}).
\par
We wish to stress that in all our comparisons with the experimental data,
including Fig.8, we include the SC, which have been
calculated in the previous section. This  means that the formulae for
the cross sections, obtained in section 2, are multiplied
by  the damping factor
$D^2_L$ or $D^2_T$, calculated in section 3.
\par
We now concentrate on the $x_P$ dependence of $F^{DD(3)}_2$.
Based on the Ingelman-Schlein
hypothesis of the Pomeron structure function\cite{IS}, $F^{DD(3)}_2$
can be written in
a factorised form
\beq \label{N1}
F^{DD(3)}_2\,\,=\,\,f(x_P)\,\,F^P_2 (Q^2,\b)\,\,
\eeq
where $F^P_2$ denotes the Pomeron structure function and $f(x_P)$
the Pomeron flux factor. If the Pomeron is a simple Regge pole then
$f(x_P) \,\,\propto\,\,
(\frac{1}{x_P})^{1+n}$. $n$ is
intimately related to the intercept of the Pomeron trajectory:
$\alpha_P\,\,=\,\,\alpha_p(0)\,\,+\,\,\alpha_P'|t|$, where
$\alpha_P(0)\,=\,1\,+\,\epsilon$. Thus: $n\,=\,
2\,\alpha_P(0)\,-\,2\,=\,2\,\epsilon$.
\par
Our approach is orthogonal to the IS one. We do not expect
\eq{N1} to scale, and if we use the parametrisation suggested
in \eq{N1}, and fit the $x_P$
dependence assuming the form
\beq \label{N2}
F^{DD(3)}_2\,\,=\,\,(\frac{1}{x_P})^{n}\,\,F^P_2(Q^2,\b)
\eeq
we expect  $n$ to depend on both $\b$ and $Q^2$.
\par
We present the calculated values of
$n(Q^2,\b)$ in Fig.9a. We would
like to point out that the value of $n$ is much
bigger ($n\,\approx\,0.55$) than the experimentally fitted value of
($n\,\approx\, 0.2$)
(see Ref.\cite{H1DDNEW}). The difference is due to the
steep behaviour of the gluon
density function in the GRV parametrisation. Notice, however, that
the new ZEUS data\cite{ZEUSNEW} gives a value of $n\,\approx\,$ 0.4
which is much closer to  our result.
Direct comparison with
the experimental data given in Fig.9a, shows that our calculation is able
to describe the experimental data at $x_P \,\approx \,10^{-3}$ but it
substantially overshoots the experimental data
for smaller values of $x_P$.  We think this behaviour is
an artifact of the GRV
parametrisation in the leading order of pQCD. A simple ad hoc procedure
to improve our results would be to take GRV in the next to leading order.
In our opinion such a procedure is not self consistent unless the
entire calculation is done in a higher order.
One can see from Fig.9b, where the ratio of the leading order GRV
gluon density to the next to leading order distribution
($ x_P G(Q^2,x_P)^{LO}/ x_P G(Q^2,x_P)^{NLO}$) is plotted
that the use of $x_P G(Q^2,x_P)^{NLO}$ instead of $x_P G(Q^2,x_P)^{LO}$
suppresses the value of the diffractive cross section at small $x_P$.
It is interesting to note that the effective power $n$ reduces
then to a value
$n\,\,\approx\,\,0.3$ which is closer to the experimental value
suggested by the new ZEUS data\cite{ZEUSNEW}.
The second source of suppression at low $x_P$ comes from
the energy behaviour of the slope
$B$, where $B\,=\,B_0\,\,+\,\,2\,\alpha'_P\,\ln(1/x_P)$.
Note that for a Regge soft Pomeron we have
$\alpha_P(t)\,=\,1 \,+\,\epsilon \,+\,\alpha'_P \,t$.
\par
Regardless of the above, the fact that pQCD calculations
predict a larger
cross section than the experimental one, lends support to our statement
in the previous subsection that at small distances, where we can
trust the pQCD approach, the calculation of the
diffractive production of small mass, appears to work.
How the $x_P$ dependence is affected by the different parametrizations of
the structure function is still an open question. We plan on clarifying
this in a future publication.

\subsection{The diffractive structure function $F^{DD(2)}_2$}
In Fig.10 we display our calculated values for $F^{DD(2)}_2$ defined as
\beq \label{N3}
F^{DD(2)}_2\,\,=\,\,\int^{(x_P)_{max}}_{(x_P)_{min}}\,\,d x_P
\,\,F^{DD(3)}_2
\eeq
where the values of $(x_P)_{min}$ and $(x_P)_{max}$ are taken to
have the same values as in the relevant experiments.
In Fig.11 we compare the calculated $F^{DD(2)}_2$ values
with the relevant H1 data\cite{H1DDNEW}.
>From these figures we are able to conclude:
\newline
1. The pQCD calculation provides a fair description of the experimental
data.
\newline
2. The contribution of the longitudinal polarised photon is important
at large  $\b \,>\,0.7$.
\newline
3. The process with an emission of an extra gluon gives rise to a
sizeble contribution for
$\b\,<\,0.4$ and should be taken into account in a more consistent
manner by
solving the evolution equations for DD processes (see the next section).
\newline
4. For $\b \,>\,0.4$, the emission of an  extra gluon provides only a
small
contribution, so one can attempt to extract the gluon density by
measuring $F^{DD(2)}_2$ at large
$\b$.

\subsection{The transverse momentum spectra}
\par
Our formulae ,see eqs. (23),(28),(36),(37) and (41), can be used
for a more detailed
analysis of the DD events. In particular, we can describe the transverse
momentum spectra of the parton jets in DD. Basically, these spectra
are given by the integrand of our formulae, as the $k_{\perp}$ which
appears is just the transverse momentum of the produced jet.
\par
In Fig.12, the ratio
\beq \label{N4}
R\,\,=\,\,\frac{F^{DD(2)}_2(k^2_{\perp}\,>\,k^2_0)}{F^{DD(2)}_2}
\eeq
is presented with different values of $k^2_0$. The physical meaning of
this ratio is that it indicates the
fraction of the all DD events possessing  transverse momenta larger
than $k_0$. One can see from Fig.12 that we expect a fairly large
fraction of the events with big transverse momenta, this  is in
agreeement with the new H1 data\cite{H1PT}.
\par
In Fig.13 we plot  the calculated value
$$
 x_P F^{DD(4)}_2\,\,=\,\,\frac{Q^2}{ 4 \,\pi^2\,B\,\alpha_{em}}
\,\frac{ x_P\,d^3 \,\s}{ d x_P d t d k^2_{\perp}}\,|_{t = 0}
$$
where $k_{\perp}$ denotes  the transverse momentum of the jet.
The comparison with the experimental data is also quite good,
at least, the $k_{\perp}$ distribution reproduces the main qualitative
feature of the experimental data \cite{H1PT}, namely, smoother behaviour
than $\frac{1}{k^4_{\perp}}$ at small values of $k^2_{\perp}$, and
$\frac{1}{k^4_{\perp}}$ behaviour
for larger $k_{\perp}$. On the other hand, we predict a sharp drop
at large $k_{\perp}$, which has not been seen in the data as yet.
Perhaps, the emission of two gluons starts to be important at such large
values of transverse momenta.

\subsection{Matching with diffractive vector meson electroproduction}
\par
Vector meson diffractive electroproduction
corresponds to the kinematic region of $\b\,
\rightarrow\,1$, namely
\beq \label{V1}
\b\,\,=\,\,\frac{Q^2}{Q^2\,+\,M^2_V}\,\,\rightarrow\,1\,\,-\,\,\frac
{M^2_V}{Q^2}
\,\,
\eeq
where $M^2_V$ is the mass of the produced vector meson. In this
kinematic region we can rewrite \eq{22} and \eq{62} in the form
\beq \label{V2}
M^2 \,\frac{d \s^L}{d M^2 d t}\,|_{t = 0}\,\,=\,\,
\eeq
$$
\frac{4\,\pi^2 \,\alpha_{em}\,\as^2}{3}\,\sum_F\,Z^2_F\,\,\frac{M^2_V}
{Q^6}\,
\int^{\frac{M^2_V}{4}}_{Q^2_{min}}\,\,\frac{d \,k^2_{\perp}}{k^2_{\perp}}
\,
\frac{1}{\sqrt{1\,-\,\frac{4 k^2_{\perp}}{M^2_V}}}\,\(\,x_P\,
G\(\frac{k^2 \,Q^2}{M^2_V},x_P\)\,\)^2
$$
\beq \label{V3}
M^2 \,\frac{d \s^T}{d M^2 d t}\,|_{t = 0}\,\,=\,\,
\eeq
$$
\frac{4\,\pi^2 \,\alpha_{em}\,\as^2}{3}\,\sum_F\,Z^2_F\,\,\frac{M^3_V}
{Q^8}\,
\int^{\frac{M^2_V}{4}}_{Q^2_{min}}\,\,\frac{d \,k^2_{\perp}}{k^4_{\perp}}
\,
\frac{1}{\sqrt{1\,-\,\frac{4 k^2_{\perp}}{M^2_V}}}\,
\{ 1\,-\,2\,\frac{k^2_{\perp}}{M^2_V}\,\}\,\(\,x_P\,
G\(\frac{k^2 \,Q^2}{M^2_V},x_P\)\)^2
$$
These formulae give the cross sections for production of all hadrons with
mass $M_V$, and reproduce the main features of the exclusive vector
meson production\cite{RY}\cite{FIVE}\cite{GLMVEC}. It is interesting to
calculate the cross sections in the region of $M^2_V\,=\,m^2_{\rho}$
where we
do not expect any other mesons, besides the $\rho$-meson to be produced.
In the double log approximation of pQCD,  we can put the
argument of the gluon density equal to $Q^2$. Evaluating the
remaining integrals we  obtain
\beq \label{V4}
\frac{\s^L}{\s^T}\,\,=\,\,
\frac{Q^2 \,Q^2_{min}\,ln\frac{M^2_V}{4\,Q^2_{min}}}{M^4_V}\,\,
\eeq
The value we take for $Q^2_{min}$
depends on how strong our belief in hadron-parton
duality is. If we believe that the hadron-parton duality can be used
at large distances,
we can
choose the low limit of integration from the condition that the
argument of the gluon density in \eq{V2} and in \eq{V3} is bigger
than the intitial virtuality in the GRV parametrization, i.e.
$Q^2_0 = Q^2_{min} Q^2/M^2_V$ or $Q^2_{min}\,=\,Q^2_0 M^2_V/Q^2$.
For such a value of $Q^2_{min}$ we have $\s^L/\s^T\,= \frac{Q^2_0}{M^2_V}
 \,\ln \frac{4\,Q^2}{Q^2_0}$.
\par
Experience gained from $e^+ e-$ annihilation indicates that $Q^2_{min}$
could be very
small, about $m^2_{\pi}$\cite{SCALE}. In this case, we still have
the $Q^2$ rise of the ratio, but the coefficient is very small,
$\s^L/\s^T
\,=\,0.125 Q^2$.
In Fig.14 we plot $\s^L$ and $\s^T$, using  \eq{V2} and \eq{V3} with a
cutoff $Q^2_{min} = m^2_{\pi}$.
\par
Concluding this section, we would like to reiterate our
claim that the simple pQCD calculation provides a
reliable basis for the discussion of the origin of DD in DIS.
It also suggests a new way to determine the
gluon density.
This new measurement can be done in two unambigous ways:
\newline
1. the measurement of $F^{DD(2)}_2$ with $\b\,>\,$ 0.7, allows us to
extract the longitudinal polarised photon structure function which is
only sensitive to small
distances of the order $r_{\perp}\,\,\approx\,\,1/Q$.
\newline
2. the measurement of $F^{DD(2)}_2$ with $k_{\perp}\,>\,k_0\,\,\geq\,
1\,GeV^2$.
In this case only distances $r_{\perp}\,\leq\,1/k_0$ contribute to our
formulae, and we have reliable pQCD predictions for the fraction of the
DD events.

\section{Evolution equations for DD}
\par
In this section we discuss the evolution equations for the DD structure
functions that have been introduced in the previous section.
The evolution equations for the transverse
polarised photon have been proposed and derived in
Ref.\cite{LW}, we will comment on them later. We
first discuss the evolution equations for the
longitudinal polarised photon, which have not been
formulated previously. These provide an interesting
theoretical insight into the structure of the evolution of
exclusive processes in QCD.

\subsection{Evolution equation for DD of a
{\bf $\bar q q$}-pair production by a
longitudinal photon}
\par
We start with the evolution equation
for the simple case of $\bar q q$
production (see \eq{22}). It is useful to introduce
the DD structure function\cite{LW}
absorbing all extra powers of $Q^2$ in the definition
\beq \label{27}
F^{DD}_L \,\,=\,\,
 x_P\,\frac{d \s}{ d x_P\,d t } \mid_{t = 0 }\,\,\,\,\{\,\,4\,\pi \,
\alpha_{em}\,\,\sum Z^2_F
\,\,\frac{1}{Q^4}\,\,\}^{-1}\,\,=
\eeq
$$
\frac{\as^2}{N_C}\,\,\b^2(1 - \b)\,(1 - 2 \b)^2
\,\,\int^{\frac{M^2}{4}}_{Q^2_0}\,\,\frac{ d\, k^2_{\perp}}{k^2_{\perp}}
\,\(\,x_P,\,G\(\,\frac{k^2_{\perp}}{\,(1 - \b)},x_P\,\)\,\)^2
$$
Taking the derivative with respect to $\ln Q^2$, we obtain the evolution
equation
\beq \label{28}
\frac{ \pa F^{DD}_L (Q^2,x_P,\b)}{\pa \ln Q^2}\,\,=\,\,
\frac{\as^2}{N_C}\,\,\b^2(1 - \b)\,(1 - 2 \b)^2
\,\,
\(\,x_P\,G\(\,\frac{Q^2}{4\,\b},x_P\,\)\,\)^2
\eeq
\par
We first investigate \eq{28} by considering the contribution of the
diffractive processes to the deep inelastic structure function (i.e. to
the total cross section). We integrate \eq{28} as well
as \eq{27} over all values of $1\,\,>\,\,x_P \,\,>\,\,x_B$.
The  solution of the evolution equation for the integrated
$F^{DD}_L (Q^2,\b)$
$(F^{DD}_{L;int}\,\,=\,\,\int^1_{x_B}\,d\,x_P \,F^{DD}_L(Q^2,x_P,\b))$
can be obtained by transforming to the moment representation
with respect to $\ln (1 / x_B)$
(see \eq{2}). This has the form
\beq \label{29}
\frac{d f^{DD}_{L;int}(Q^2,\o)}{d \ln Q^2}\,\,=\,\,\as^2\, f^{G^2}_
{L;int}(Q^2,\o)
\eeq
where we denote by $f^{G^2} (Q^2,\o)$ the moment of the right hand side
of \eq{28} integrated over $x_P$. Our calculation is done in
the double log approximation where
we know that\cite{GLR}\cite{HT}
\beq \label{30}
f^{G^2}_{L;int}(Q^2,\o)\,\,=\,\,f^{G^2} (Q^2_0,\o) \,\,e^{\g_2(\o)\,\ln
 \frac{Q^2}{Q^2_0}}
\eeq
and
\beq \label{G2O}
\g_2(\o)\,=\,\frac{4\,N_C \as}{\pi}\,\,\frac{1}{\o}\,\,\gg\,\,\g_
1(\o)\,=\,\g(\o)\,=\,\frac{N_C \as}{\pi}\,\,\frac{1}{\o}
\eeq
The solution of \eq{29} is
\beq \label{31}
f^{DD}_{L;int}(Q^2,\o)\,\,=\,\,\as \frac{f^{G^2}(Q^2_0,\o)}{\g_2(\o)} \,
\,e^{\g_2(\o)
\,\ln \frac{Q^2}{Q^2_0}}
\eeq
For the deep inelastic structure function,
the process of DD initiated by a longitudinal
polarized photon contributes to the twist four term in the Wilson
Operetor Product Expansion (note the additional factor of
$\frac{1}{Q^2}$ in \eq{22}). We claim that the anomalous dimension of
the twist four operator is equal to $\g_2$. It should be stressed that
the diffractive cross section does not describe the complete twist four
contribution to the structure function. The additional gluon
emission which increases the value of the
total anomalous dimension\cite{HT} does not contribute to the
diffraction into small masses.

\subsection{Evolution of the DD structure function (greneral
considerations)}
\par
In the general case we can rewrite \eq{27} in the form
\beq \label{32}
F^{DD}_L\,\,=\,\,
\frac{\as^2}{N_C}\,\int^{Q^2(1 - \b)}_{Q^2_0}\,\,\frac{ d\, k^2_{\perp}}
{k^2_{\perp}}\,\,\int^1_{\b} d z \frac{\b}{z}\,
\Sigma^S_S(\frac{ \b}{z},\ln \frac{Q^2}{k^2_{\perp}})\,\,\Phi^F_{P;L}
\(x_P\,G\(\,\frac{k^2_{\perp}}{\,(1 - \b)}, x_P\,\)\,\)^2
\eeq
where
$\S $ denotes the singlet structure function
of the quark-antiquark pair
($q(x,Q^2) + \bar{q}( x,Q^2)$) and
the Pomeron splitting function $\Phi^F_L$ has been defined above
(see \eq{PHIL}). It should be stressed that only $\Sigma^S_S$ enters
the evolution equation in the leading order of pQCD.
\par
Differentiating the above equation we obtain the evolution equations
for the monents in the form
\beq \label{33}
\frac{d f^{DD;S}_{L;int}(Q^2,\o)}{d \ln Q^2}\,\,=
\eeq
$$
\, P^S_S(\o)
\,\,\,\,f^{DD;S}_{L;int}(Q^2,\o)\,\,+\,\,P^S_G(\o)\,\,\,
\,f^{DD;G}_{L;int}(Q^2,\o)\,\,+\,\,
\frac{\as^2}{N_C}\, f^{G^2}(Q^2,\o)\,\,
$$
where $P$ are the kernels of the DGLAP evolution equations. We obtain
the ordinary DGLAP evolution equations with an inhomogenious term.
This equation has the solution
\beq \label{34}
f^{DD}_{L;int}(Q^2,\o)\,\,=
\eeq
$$
\,\frac{\as^2}{N_C} \,\,
f^{G^2}(Q^2_0,\o)\,\{\,\frac{1}{ \g_2(\o) \,-\,\g^{+}(\o)} \,+\,\frac{1}
{g_2(\o)
 \,-\,\g^{-} (\o)}\,\}\,
 e^{\g_2(\o)\ln \frac{Q^2}{Q^2_0}}
$$
$$
\,\,+\,\,f^{DD}_{L;int}(Q^2_0,\o)
\,\,\{\,\frac{ -\,\g^{-}(\o)}{\g^{+}(\o) \,-\,\g^{-}(\o)}
e^{\g^{+}(\o)\,\ln \frac{Q^2}{Q^2_0}}\,\,
+\,\,
\frac{ \g^{+}(\o)}{\g^{+}(\o)\,-\,\g^{-}(\o)}\,
e^{\g^{-}(\o)\,\ln \frac{Q^2}{Q^2_0}}\,\}
$$
where the functions $f^{DD}_{L;int}(Q^2_0,\o)$ and $f^{G^2}(Q^2_0,\o)$
denote the
intial conditions for the evolution equations, and describes the large
distance physics. It is obvious that the second term of \eq{34} is
the solution of the DGLAP equations for fixed $\as$, which we
assumed in order to
simplify the solution. One can find this form of the solution
in Refs.\cite{MUBOOK}\cite{EKL}
and in the references contained therein.
$\g^{+,-}$ denotes the eigenvalues of the
anomalous dimensions matrix
and in the leading order of pQCD they are equal to
\beq \label{AD1}
\g^{\pm} \,\,=\,\,\frac{1}{2}\,\{ P^S_S  \,\,+\,\,P^G_G\,\,\pm\,\,
\sqrt{ ( \,P^S_S\,\,-\,\,P^G_G\,)^2 \,\,+\,\,4 P^S_G P^G_S}\,\}
\eeq
where
$$
P^S_S\,=
\,\frac{C_F \as}{2 \pi}\,\{\frac{3}{2}\,-\,\frac{1}{\o + 1} \,-\,
\frac{1}{\o +2} \,- \,2 \,[ \psi(\o + 1) \,-\,\psi(1)\,]\,\}\,\,
$$
$$
P^G_G\,=\,\frac{\as}{2 \pi}\,\{ 2 N_C\, [\,\frac{11}{12} \,+\,\frac{1}
{\o}\,-
\,\frac{2}{\o + 1} \,+\,\frac{1}{\o + 2} \,-\,\frac{1}{\o + 3} \,-
\psi(\o + 1)
+ \psi(1) \,] \,-\,\frac{2 T_R N_F}{3}\,\}
\,\,
$$
\beq \label{AD2}
P^G_S\,=\,\frac{ C_F\as}{2\,\pi}\,\,\{\,\frac{2}{\o}\,-\,\frac{2}{\o + 1}
\,
+\,\frac{1}{\o + 2}\,\}
\eeq
$$
P^S_G\,=\,\frac{2 T_R N_F\as}{2\,\pi}\,\,\{\,\frac{2}{\o + 3}\,-
\,\frac{2}{\o + 2}\,+\,\frac{1}{\o + 1}\,\}\,\,
$$
It should be stressed that in pQCD only $\g^{+}$ has a singularity in
the second term at small
$\o$,  and $\g^{+} \,\,\ll\,\,g_2(\o)$
at $\o \,\ra\,0$.
This means that only the first term
survives at small $x$.
\par
We do not claim that this result is valid at small virtualities, where
pQCD arguments are not applicable.
The value of the DD structure function at a small vituality
of the photon ($Q^2 \,=\, Q^2_0$) which can be measured experimentally,
defines only the specific sum of singularities, namely
\beq  \label{35}
f^{DD;S}_{L;int}(Q^2_0,x_B)\,\,=
\eeq
$$
\,\frac{1}{2\,\pi\,i}\,\int_C\,d \o
\,\(\,\frac{\as^2}{N_C}\,f^{G^2} (Q^2_0,\o)\,[\frac{1}{\g_2(\o) -
\g^{+} (\o)} \,+\,\frac{1}{\g_2(\o) -
\g^{-} (\o)}] \,\,+\,\,f^{DD;S}_{L} (Q^2_0,\o)\,\)
$$
The fact that these two terms have different anomalous dimensions,
suggests that one might be able
to use their $x_P$ dependence in order to distinguish between them.
\par
The second term is
proportional to
$ (\,x_P G(Q^2_0,x_P))^2$ for the unintegrated $F^{DD}$,
but there is no theoretical argument why
such a behaviour should be valid for a small value of $Q_0$.
Phenomenologically,
we associate these two contributions with the "soft" and "hard"
Pomeron.
This terminology reflects our hope that their
$x_P$ behaviour is different. Assuming that their behaviour can be
approximately parameterized as $ x^{2\,\epsilon}_P$, we expect for the
"soft"
(second) term  $\epsilon_{soft}\,\,\sim\,\,$0.08 - 0.1, while for
the "hard" one (first term in \eq{34}) $\epsilon_{hard}\,\,\sim\,$
0.2-0.5.
We wish to stress that the problem of separating these two
contributions is beyond the scope  of pQCD, and has to be tackled using
a nonperturbative approach. As a guide to phenomenological applications
we would like to mention that the Ingelman and Schlein approach\cite{IS}
based on the
Pomeron structure function, means that only the "soft" contribution
survives, while the "hard" Pomeron approach deals only with the first
tem in \eq{34}.
Returning to $F^{DD}$
(see \eq{32}), we can write the evolution equation at fixed $x_P$
\beq \label{36}
\frac{ \pa \,\,F^{DD;S}_{L}(Q^2,x_P,\b)}{\pa\,\,\ln Q^2}\,\,=
\int^1_{\b} d z \,
P^S_S(z)\,\,F^{DD;S}_L(Q^2,x_P,\frac{\b}{z})\,\,+
\eeq
$$
\int^1_{\b} d z \,
P^S_G(z)\,\,F^{DD;G}_L(Q^2,x_P,\frac{\b}{z})\,\,+
\,\,\frac{\as^2}{N_c}\,
\, \b^2 \, ( 1 - 2 \b)^2 \,\,\(x_P \,G\(\frac{Q^2}{ 4 \b},x_P\)\)^2
$$
In moment space with respect to $\ln (1/ \b)$, the equation reduces
to the form
\beq \label{37}
\frac{ d \,f^{DD;S}_L (Q^2,x_P,\o)}{d \ln Q^2}\,\,=
\eeq
$$
\,\,
P^S_S(\o) \,f^{DD;S}_L (Q^2,x_P,\o)\,\,+\,\,P^S_G(\o) \,f^{DD;G}_L
(Q^2,x_P,\o)
\,\,+\,\,T(\o)\,\(\,x_P \,G(Q^2,x_P)\,\)^2\,\,
$$
where
\beq \label{37.1}
T^F_{G;L}(\o) \,=\,\frac{1}{ \o + 2} \,-\,\frac{4}{ \o + 3}\,
+\,\frac{4}{\o + 4}
\,\ra|_{\o \ra 0}\,\,\, \frac{1}{6}
\eeq
The general solution of the above equation is
\beq  \label{38}
f^{DD}_{L;int}(Q^2,\o)\,\,=
\eeq
$$
\,\frac{\as^2}{N_C} \,\,
\{\,f^{DD}_{L;int}(Q^2_0,\o)
\,\,\{\,\frac{ -\,\g^{-}(\o)}{\g^{+}(\o) \,-\,\g^{-}(\o)}
\,e^{\g^{+}(\o)\,\ln \frac{Q^2}{Q^2_0}}\,\,+\,\,
\frac{ \g^{+}(\o)}{\g^{+}(\o) \,-\,\g^{-}(\o)}
\,e^{\g^{-}(\o)\,\ln \frac{Q^2}{Q^2_0}}\,\}
$$
$$
+\,\,
[\,\frac{\g_2(\o)}{\g_2(\o)\,-\,\g^{+}(\o)}\,+\,\frac{\g_2(\o)}{\g_2(\o)
\,-\,\g^{-} (\o)}\,]
\int^{Q^2}_{Q^2_0} \,\frac{d Q'^2}{Q'^2}\,T(\o)\,\,\(x_P \,G(Q^2,x_P)\,
\)^2
$$
$$
=\,\,f^{DD}_{SOFT} + f^{DD}_{HARD}\,\,
$$
The first term $F^{DD}_{SOFT}$ is associated with soft diffraction, and
it requires the phenomenological input $f^{DD}(\o,x_P,Q^2_0)$.
We choose for this input
\beq \label{EV1}
  f^{DD}(Q^2_0,x_P,\o)\,\,=\,\,T(\o)
\,\(x_P \,G(Q'^2,x_P)\,\)^2
\eeq
We still need to fix the value of $Q_0$ in our
initial condition. There is no reason to assume the same value of $Q_0$
in DD and DIS. We  choose, for DD, a value of $Q_0^{2}$ which is
approximately $1- 2 \,GeV^2$. In doing so we can use pQCD to evaluate the
two terms appearing in \eq{38}, and this leads to \eq{EV1} for the first
term.
\par
In the $\b$-representation the intial condition is simply
\beq \label{EV2}
F^{DD}_{L}(Q^2_0,x_P,\b)\,\,=\,\,\frac{\as^2}{N_C}\b^2\,(1\, - \,2\b)^2
\,\(\,x_P\,G\(\frac{Q^2_0}{4 \b},x_P\)\,\)^2
\eeq
This is plotted in Fig.15 for different values of $x_P$.

\subsection{Evolution equation for transverse polarized photon initiated
DD}
The evolution equations for a transverse polarised photon have been
given in Ref.\cite{LW}. Introducing the densities $q_P (Q^2,x)$ and
$G_P(Q^2,x)$ which denote the quark and gluon densities in the Pomeron,
we can rewrite
the evolution equations in the form (see ref.\cite{LW} for details):
\beq \label{EV3}
\frac{\pa\, \S_P (Q^2,x)}{\pa \ln Q^2}\,\,=
\eeq
$$
\,\,\int^1_{\b}\,\frac{d z}{z}
\,{P^S_S(z) \,\S_P(Q^2,\frac{\b}{z})\,\,+
\,\,P^S_G(z)\,G_P(Q^2,\frac{\b}{z})\,
}\,\,+\,\,\frac{1}{Q^2}\,\Phi^q_{P;T} (\b) \,\(\,x_P\,G(Q^2_0,x_P)\)^2
$$
$$
\frac{\pa\, G_P (Q^2,x)}{\pa \ln Q^2}\,\,=
\,\,\int^1_{\b}\,\frac{d z}{z}
$$
$$
\,\{P^G_S(z) \,\S_P\(( 1 - \b)Q^2,\frac{\b}{z}\)\,\,
+\,\,P^G_G(z)\,G_P\((1 - \b) Q^2,\frac{\b}{z}\)\,
\}\,\,
$$
$$
+\,\,\frac{1}{Q^2}\,\Phi^G_{P;T} (\b) \,\(\,x_P\,G(Q^2_0,x_P)\,\)^2\,\,
$$
where $\S\,= q (Q^2,x)\,+\,\bar{q}(Q^2,x)$.
\par
On inspection we note
that the structure of the equations is the same as for the
longitudinal polarised photon, \eq{EV3}, and is a
normal DGLAP evolution equation
with an inhomogenious term. $P$ are the kernels of the DGLAP
equations while $\Phi $ denotes the Pomeron splitting functions
\beq \label{EV4}
\Phi^S_{P;T}\,\,=\,\,\frac{\as}{N_C}\,\,z^2\,(1 - z)^2\,\,
\eeq
$$
\Phi^G_{P;T}\,\,=\,\,\frac{\as N^2_C}{2 (N^2_C - 1)}
\,\frac{1}{z}\,\,(1 - z )^2 \,( 1 + 2 z )^2\,\,
$$
The set of \eq{EV3} can be solved in the moment representation using
the standard techniques (see Refs.\cite{MUBOOK}\cite{EKL} and references
therein). The main difference in the comparison with the lomgitudinal
photon, is the fact that the inhomogenious term has an extra $1/Q^2$
suppression, provided we choose the appropriate
initial condition for \eq{EV3}. The contribution of this term
to the evolution equation is consequently small.
To understand the structure of the result we write the solution of
\eq{EV3} in the region of small $\b$ ($\o \,\ra\,0$ ),
where the emission of extra gluons
and quarks could be important (as have been seen in section 3).
In the region of small $\o$ we can neglect $\g^{-}$ (see for example
Ref.\cite{EKL}) and the solution has a simple form
\beq \label{EV5}
\S_P(Q^2,\o)\,\,=
\eeq
$$
\,\{\,\S_P(Q^2_0,\o)\,\,\frac{\g^{-}(\o)}{\g^{+}(\o) \,-
\,\g^{-}(\o)}\,\,+\,\,G_P (Q^2_0,\o)\frac{2 N_F P^S_G (\o)}{\g^{+}(\o)
\,-
\,\g^{-}(\o)}\,\}\,\,
e^{\g^{+}(\o)\,\ln (Q^2/Q^2_0)}\,\,
$$
$$
+\,
\,\int^{Q^2}_{Q^2_0}\,\,
\frac{d Q'^2}{Q'^4}
\frac{\g_2(\o)\,[\Phi^S_{P;T} (\o)\,+\,\Phi^G_P(\o)]}
{\ti{g}_2(\o) \,-\,\g^{+}(\o)}
\(x_P\,G(Q'^2,x_P)\,\)^2
$$
where $\ti{g}_2(\o)\,\ra\,\frac{4 N_c \,\as}{\pi \o} - 1 $
at $\o\,\ra\,0$, and $\Phi_P (\o)$ is the $\o$ image of the Pomeron
splitting functions of \eq{EV4}. They are of the form
\beq \label{EV6}
\Phi^S_{P;T}\,\,=
\,\frac{\as}{N_C}\,\{\,\frac{1}{\o + 2} - \frac{2}{\o + 3}
+ \frac{1}{\o + 4}\,\}\,
\eeq
$$
\Phi^G_{P;T}\,\,=
\,\,\frac{\as N^2_C}{2 (N^2_C - 1)}\,
\{\frac{1}{\o}\,+\,\frac{2}
{\o + 1}\,-\,\frac{3}{\o + 2}\,-\,\frac{3}{\o + 3}\,-\,\frac{4}{\o + 4}
\,+ \,
\frac{4}{\o + 5}\,\}\,\,
$$
To fix the two initial
moments $\S_P(Q^2_0,\o)$ and $G_P(Q^2_0,\o)$ which enter the
solution of the phenomenological term $f^{DD}_{SOFT}$ at low
$\b$, we need to choose a phenomenological input.
We use Regge phenomenology to fix this term\cite{CAPELA}
rewriting it in
the form
\beq \label{39}
f^{DD}_{SOFT}(Q^2_0,x_P,\o)\,\,=\,\,
(\frac{1}{x_P})^{2\varepsilon_{soft}}\,\,f^{DD}_{SOFT}(Q^2_0,\o)
\eeq
where $F^{DD}(Q^2_0,\o)$ is the $\o$ -image of
$G_{PRP}\,\,\b^{\Delta_R(0)} ( 1 - \b)^{n}$.
$\Delta_R (0) \,=\,\alpha_R(0) - 1 $ where $\alpha_R (0)$ is the
intercept of the
secondary Regge trajectory ($\alpha_R(0)\,\sim\,$ 0.5). The power $n$
is not defined in Regge phenomenology and we only know
that $n >$ 1. $G_{PRP} $ denotes the triple Pomeron-Regge-Pomeron
vertex\cite{CAPELA}.
The second term is the hard diffractive contribution, which is
calculated theoretically.
\par
There is an alternate way to implement the initial condition,
i.e. to use the results
of our calculations in the low mass region. We showed in our
calculations for small mass production, that the typical virtuality is
sufficiently
large. Therefore, we can choose the value of the
initial virtuality $Q_0^{2}$ for the
DD process to be about $1 - 2 \,GeV^2 $ and calculate $f^{DD}_{SOFT}$ in
pQCD. In this case
\beq \label{sig}
\S_P(Q^2_0,\o)\,=\,
\frac{1}{Q^2_0} \,\Phi^S_{P;T} (\o)\,\,\(x_P \,G(Q'^2,x_P)\,\)^2
\eeq
and
\beq \label{GP}
G_P(Q^2_0,\o) \,=\,
\Phi^G_P(\o)\,\(x_P \,G(Q'^2,x_P)\,\)^2
\eeq
This input corresponds to the initial distributions in the $\b$ -
representation
\beq \label{EV7}
\S_P(Q^2_0,x_P,\b)\,=
\,\frac{1}{Q^2_0}\,
\Phi^S_{P;T} (\b) \,\(\,x_P\,G(Q^2_0,x_P)\,\)^2
\eeq
$$
G_P(Q^2_0,x_P,\b)\,=\,\frac{1}{Q^2_0}\,
\Phi^G_{P;T} (\b) \,\(\,x_P\,G(Q^2_0,x_P)\,\)^2\,\,
$$
These distributions are plotted in Fig.16 for different values of $x_P$
at two $Q^2_0$ values ($Q^2_0 \,=\,1\, GeV^2$ and $Q^2_0 \,=\,2\,GeV^2$).
\par
We wish to emphasis two points. First, we propose to seperate between
the soft and hard diffraction processes in a way which conforms
with the  factorization theorem\cite{FACTOR} for the integrated
diffractive structure function.
Although, there is no proof that the factorization theorem
holds for the case of diffractive contribution,
in DIS\cite{HARDDD},
all known contributions which violate the factorization theorem
turn out to be small\cite{HT}\cite{LEVEIL}\cite{BAWU}.
\par
The second point concerns the value of the scale $Q_0$. There is no
reason to take the same value of $Q_0$, both in the evolution equation
and in the DIS structure function.
Depending on the choice of the scale $Q^2_0$, we can start the evolution
with very small values of $Q_0$ and use Regge phenomenology as an
initial condition for the DD structure function. Or, we can start with a
sufficiently large value of $Q_0$ in the DD processes, and reconstruct
the initial distributions using pQCD.
We followed the second alternative, and our calculations for small masses
supports this strategy.

\section{Conclusions}
\par
DD processes in DIS provide a new window to
collective phenomena in the parton cascades since they originate from
parton-parton interactions and vanish if there is no interaction
between the partons. However, DD processes possess an inherent
awkwardness in that it is
difficult to predict what is small or large in DD. Indeed, in two extreme
limits for small and for large ("black disc") parton-parton interactions,
the cross section of the DD turns out to be small. In this paper an
attempt is made to discuss the DD process within a pQCD framework.
\par
Our main results are:
\newline
1. The DD cross section originates from small distances ($r_{\perp}$
\,$<$\,0.2 fm) even for a transverse polarised photon.
This fact justifies our
use of pQCD for DD processes.
\newline
2. For $\b \,>\,$ 0.4 only production of $\bar q q $ pairs contributes
to the cross section of DD. This suggests a method
to extract the value of the gluon
density from the DD measurements in this kinematic region.
\newline
3. For $\b \,>\,$ 0.7 the longitudinal polarised photon provides the
dominant contibution to
DD processes. DD of a longitudinal photon is one of the
most promising processes
for pQCD calculations, since only small distances ($r_{\perp}\,\propto\,
1/Q$) contibute to this process. Therefore, the measurement of DD at
such large
values of $\b$ allows one to extract the value of the gluon density.
It has all the advantages of vector meson production in DD\cite
{FIVE}\cite{GLMVEC} but without any of the uncertainties
due to our
poor knowledge of the hadronic wave function\cite
{FKS}\cite{RRML}.
\newline
4. The general approach to the evolution equations for DD has been
formulated and solutions
to the evolution equations have been found. The main difference between
the evolution equations for DD
and the DGLAP evolution equations for the DIS structure functions,
is the appearance of the inhomogenious term in the DD equations. The form
of this term and its influence on the solution has been discussed.
The inhomogenious term in the DD evolution equations (eq. (103)) changes
the behaviourof $\frac{\partial \Sigma_P}{\partial \ln Q^2}$ at large
values of $\b$ adding a positive contribution. This contribution may
change our conclusion that the Pomeron gluon density peaks at
$\b \rightarrow 1$, which was based on the DGLAP evolution without the
inhomogenious term.
\newline
5. The formulae for SC have been obtained and the damping factors have
been calculated. The main result is that SC are important in the case
of DD, and the value of the cross section for DD
processes allows us to estimate
the corrections to the DIS structure function which turns out
to be fairly large, $\frac{\Delta F_2}{F_2}\,\approx\, 2\,
\frac{\sigma^{DD}}{\sigma_{tot}}$.
\par
We consider our paper as a first step in the understanding of DD as a new
source of  information about collective phenomena in DIS
in the region of small $x$. We hope this paper will give some
impetus to treat DD as a "hard" process, within the solid
theoretical framework of pQCD.
\newline
\newline
{\bf Acknowledgement:}
E. M. Levin wishes to thank LAFEX-CBPF for their kind hospitality and
CNPq (Brazil) for partial financial support. While completing this
paper, our attention was drawn to two recent papers\cite{GNZ}\cite{BLW}
which deal with some of the topics covered in this paper.
\newline
\newline

\newpage

\section*{Figure Captions.}

\begin{tabular} {l l}
   & \\
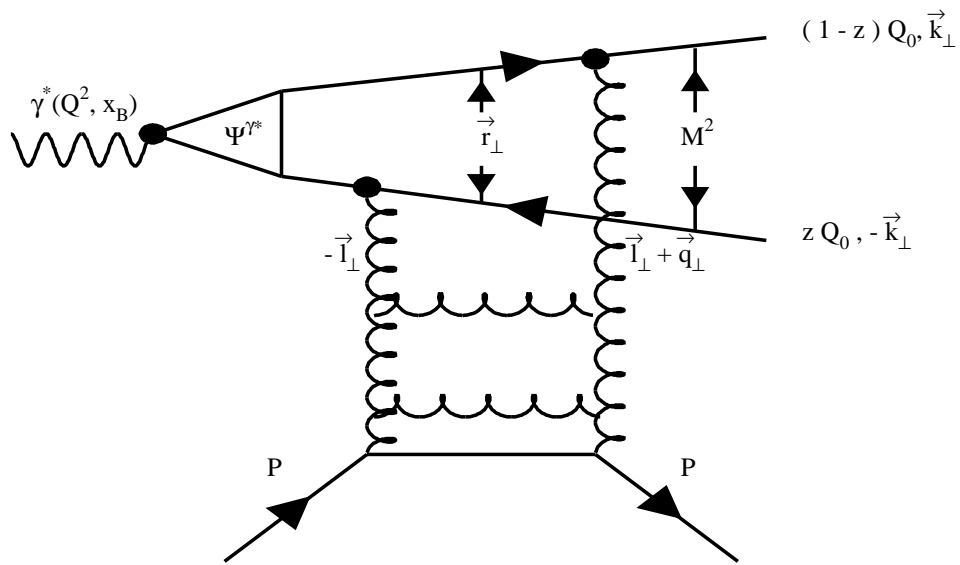
{\bf Fig.1:} & Diffractive production of a quark-antiquark pair. \\
   &   \\
{\bf Fig.2:} & Extra gluon emission in diffractive production.\\
   & \\
{\bf  Fig.3:} & The damping factor ($D_L^2$) for a longitudinal
polarised photon.
\\ & \\
{\bf  Fig.4:} & The damping factor ($D_T^2$) for a transverse
polarised photon.
\\ & \\
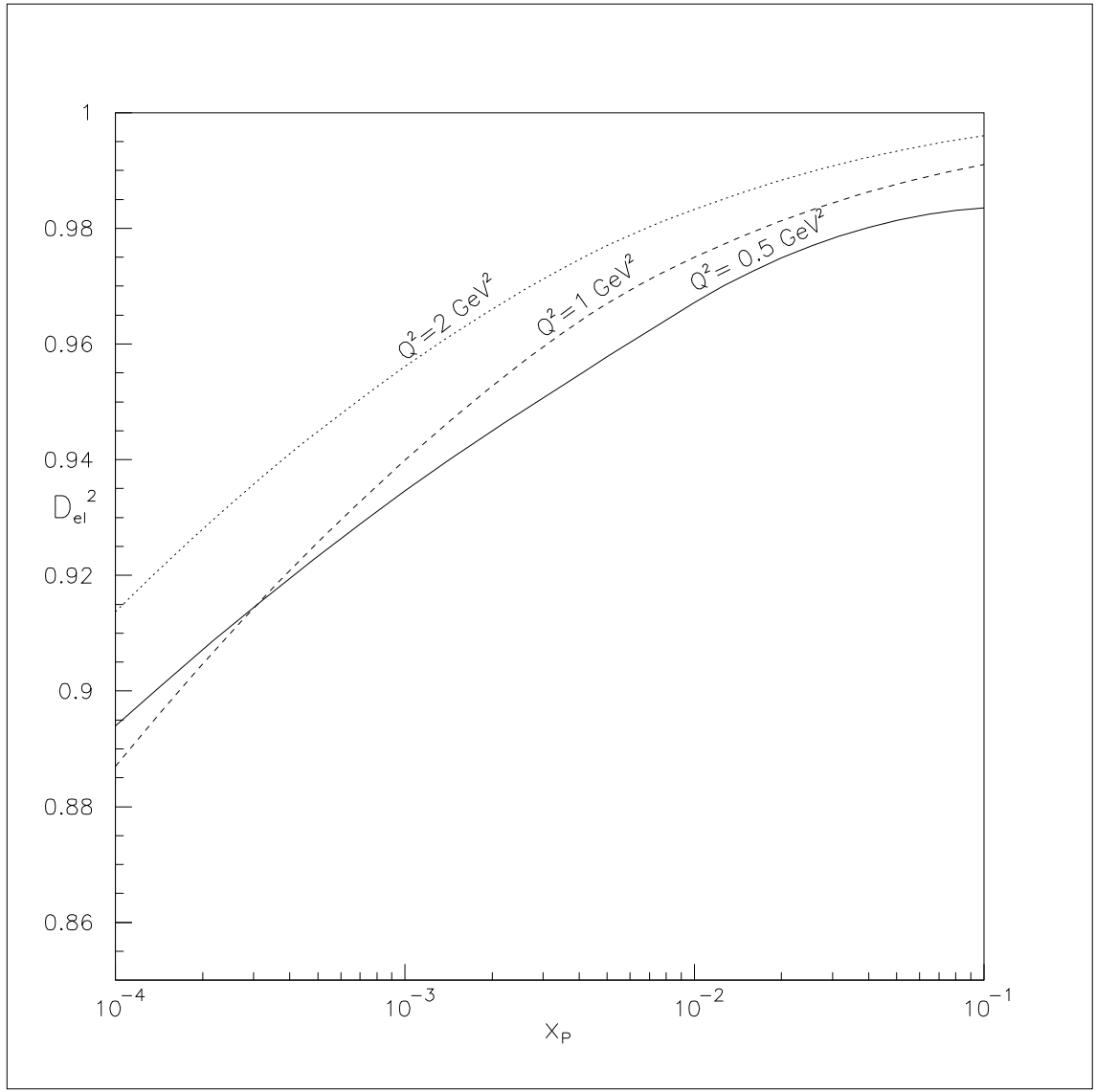
{\bf  Fig.5:} & The damping factor ($D_{el}^2$) for $F_2(Q^2,x)$.\\
 & \\
{\bf Fig.6:} & DD by longitudinal and transverse polarised photons\\
 &  at $\b$ = 0.8.\\

{\bf Fig.7:} & The integrand of eq. (28). \\
   &   \\
{\bf  Fig.8:} & $x_P\,F^{DD(3)}_2$ versus $x_P$, calculations and HERA
data. \\
  &  \\
{\bf  Fig.9:} & Effective power $n(Q^2,\b)$ and the ratio
$\frac{x_P G(Q^2,x_P)^{LO}}{x_P G(Q^2,x_P)^{NLO}}$
in the GRV paramerization.\\
 &   \\
{\bf  Fig.10:} & Different contributions to $F^{DD(2)}_2$.\\
  &  \\
{\bf  Fig.11:} &  Comparison of $F^{DD(2)}_2$  with the HERA experimental
data. \\
  &  \\
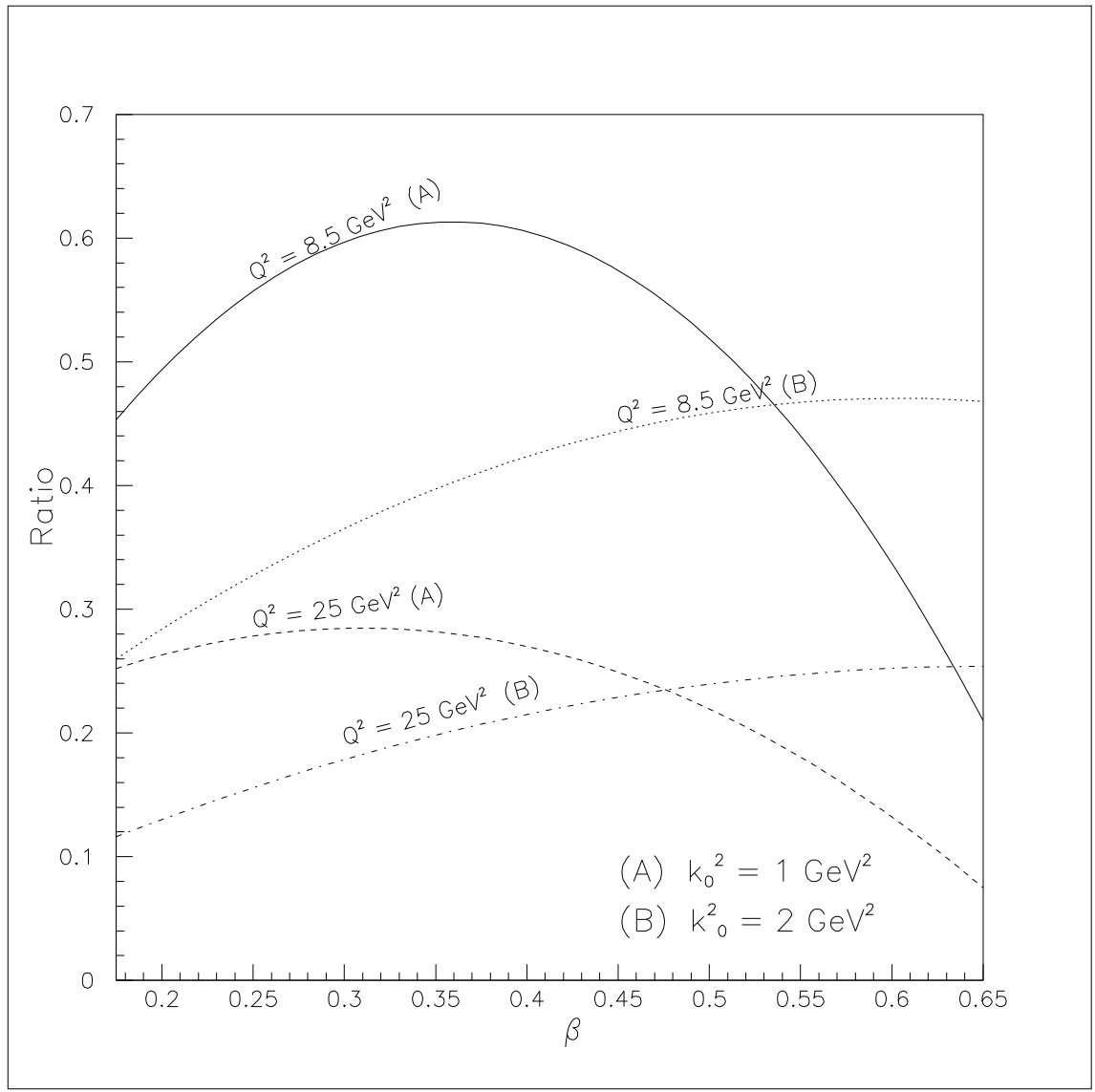
{\bf  Fig.12:} &  The ratio R (eq. (80)) for different values of $k_0$.
\\  &   \\
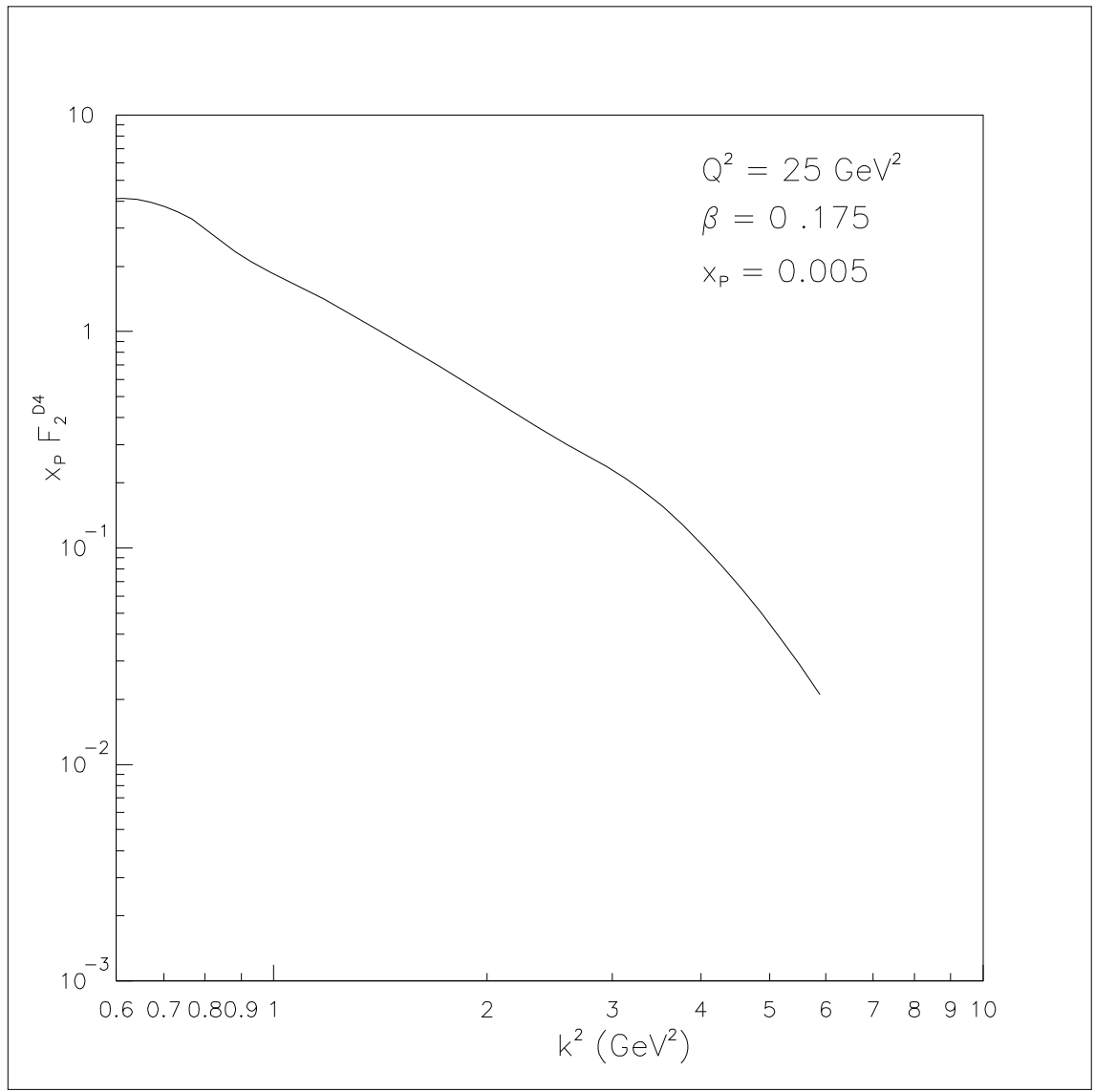
{\bf  Fig.13:} & The transverse momentum distribution for the DD jet.\\
 &  \\
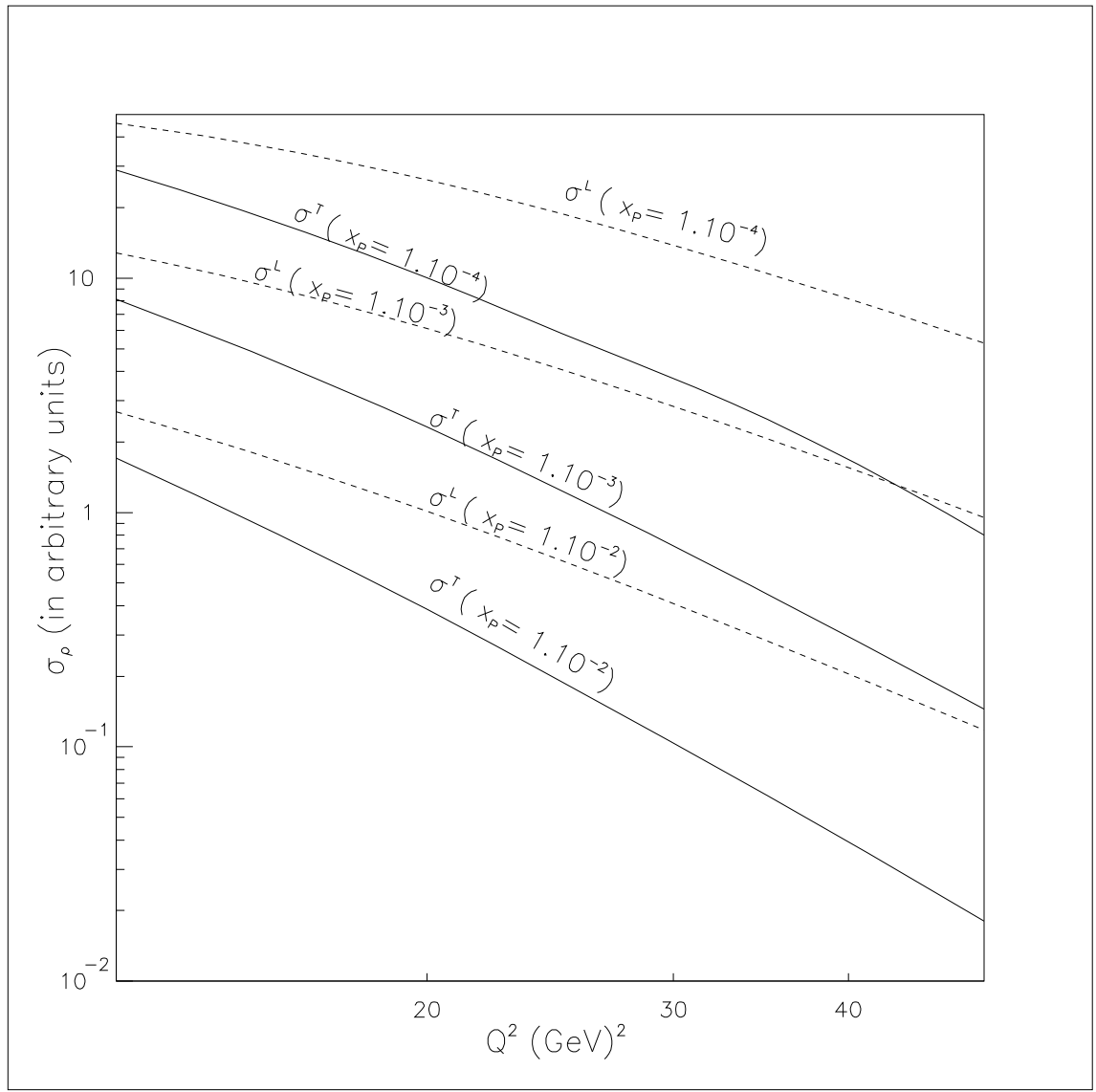
{\bf  Fig.14:} & The cross section for $\rho$ production
(in arbitrary units) for eqs. (82) and (83).\\
 &  \\
{\bf  Fig.15:} &  The initial parton distribution
for the DD evolution
equations with longitudinal \\ &  polarised photons.\\
   &  \\
{\bf  Fig.16:} &  The initial parton distributions
for the DD evolution
equations with transverse \\ &  polarised photons.\\

\end{tabular}
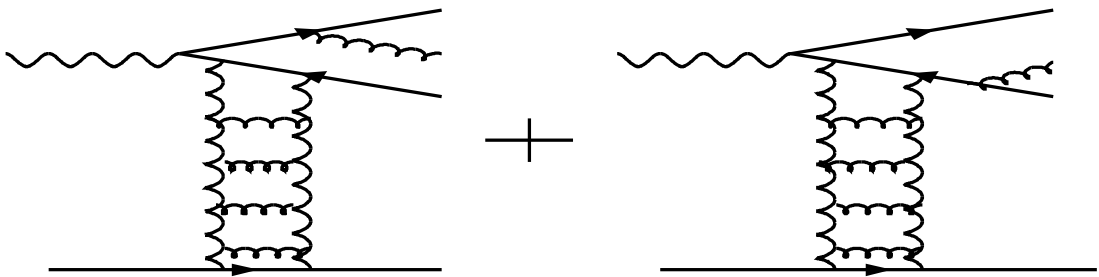
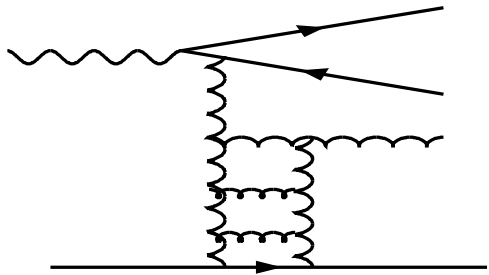
\end{document}